\documentclass[preprint,review,12pt]{elsarticle}

 \usepackage{graphics}

\usepackage{subfigure}
\usepackage{amsmath}
\usepackage{subeqnarray}
\usepackage{array}
\usepackage{arydshln}
\usepackage{gensymb}
\usepackage{marvosym}
\usepackage[numbers]{natbib}

\journal{Journal of Computational Physics.}

\begin{document}

\begin{frontmatter}


\title{A fully pressure-velocity coupled immersed boundary method based on the Lagrange multiplier approach}

\author{Yuri Feldman, Yosef Gulberg}

\address{Department of Mechanical Engineering, Ben-Gurion University of the Negev,\\
P.O. Box 653, Beer-Sheva 84105, Israel}

\begin{abstract}
A new formulation of the immersed boundary method, which facilitates accurate simulation of incompressible isothermal and natural convection flows around immersed bodies and which may be applied for accurate linear stability analysis of the flows, is presented. The method is based on the fully pressure-velocity coupled approach, implicitly satisfying the divergence-free velocity constraint with no need for an extra projection-correction step, which is a significant advantage for the computational efficiency. The method treats pressure, boundary forces, and heat sources as Lagrange multipliers, thereby implicitly providing the kinematic constraints of no-slip and the corresponding thermal boundary conditions  for immersed surfaces. Extensive verification of the developed method for both isothermal and natural convection flows is provided.
\end{abstract}


\end{frontmatter}


\section{Introduction}
\label{Intro}
Since the immersed boundary (IB) method was first introduced by Peskin \cite{peskin1972JCP}, the IB method and its modifications have become very popular numerical tools for describing the flow around moving or deformable bodies with complex surface geometry \cite{peskin2002ActaNumer, mittal2005AnnRev}. An arbitrary immersed object, whose geometry does not, in general, have to conform to the underlying spatial grid, is typically  determined by a set of Lagrangian points. At the Lagrangian points, appropriate volumetric (or surface) forces are applied to enforce the no-slip velocity boundary conditions on the body surface. These forces appear as additional unknown variables, whose values -– along with those for the pressure and velocity fields –- are obtained by solving the Navier Stokes (NS) equations. Since the location of the Lagrangian boundary points does not necessarily coincide with the underlying spatial discretization, regularization and interpolation operators must be defined to convey information to and from the body surface.

An accurate calculation of the Lagrangian forces, precisely enforcing the no-slip constraint on the surface of the immersed body, is the key issue in any IB formulation. Lagrangian forces acting on rigid bodies (as well as on bodies with a prescribed surface motion) can be treated explicitly or implicitly. Historically, explicit treatment of Lagrangian forces has gained the most attention, giving rise to the direct forcing approach, introduced by Mohd-Yusof \cite{Mohd-Yusof1997} and coauthors \cite{faldun2000JCP}, and to the immersed interface method (IIM), introduced by Lee and LeVeque \cite{lee2003SIAMSciComput} and revisited by Linnick and Fasel \cite{linnick2005JCP}. The direct forcing approach has recently been extended to thermal flow problems, see e.g. \cite{yoon2010heattransf, ren2012compufluid, ren2013IJHMT, gulb2015IJHMT}, by adding an energy equation along with the appropriate volumetric heat sources at the Lagrangian points. The direct forcing approach is not a standalone solver; rather, it may be viewed as a feature that can be easily plugged into an existing time marching solver, typically developed for the solution of NS equations on structured grids in rectangular domains. The procedure does not require any significant modifications to the existing time marching solver, which explains why the direct forcing approach is so popular. However, the direct forcing approach has a number of drawbacks. First, the no-slip condition is explicitly enforced on the intermediate non-solenoidal velocity field, whereas the divergence-free velocity field is calculated afterwards, after a projection-correction step. Second, it should be stressed that even if the NS equations are exactly solved by the projection method, resulting in a solenoidal velocity field on the Eulerian grid, the velocity interpolated to the Lagrangian points is not necessarily divergence free, which may result in a local mass leakage through the boundaries of the immersed body. Third, a pointwise local calculation of the Lagrangian forces and heat sources does not take into account their mutual interaction, which contradicts the elliptic character of the NS equations.

To improve the accuracy of the direct forcing approach, a number of techniques have been developed in the past decade. Worth mentioning here are the works of Ren at al. {\cite{ren2012compufluid, ren2013IJHMT}, who proposed an implicit evaluation of all the Lagrangian forces and heat sources by assembling them into a single system of equations. Another approach is due to Kempe at al. {\cite{kempe2012JCP, kempe2015JCP}, who introduced additional iterations to enhance Euler--Lagrange coupling, thereby providing a substantially more accurate imposition of the boundary conditions on the immersed body surface. A coupled scheme in which the momentum equations are implicitly coupled with the Lagrangian forces and heat sources and simultaneously solved as a whole system offers an alternative to the direct forcing approach. The closure of this new system is achieved by adding equations interpolating the Eulerian velocity and the temperature fields on the surface of the immersed body to enforce the prescribed boundary conditions. In this setup, the Lagrangian forces and heat sources distributed  on the fluid –- structure interface play the role of Lagrange multipliers, enforcing velocity and temperature constraints on the surface of the immersed body. A detailed explanation of this approach, together with the high-accuracy results, has been presented by Taira and Colonius \cite{taira2007JCP}, who combined the coupled IB method with a projection approach to satisfy the divergence-free and no-slip kinematic constraints. A similar idea underlies the distributed Lagrange multiplier method (DLM) of Glowinski et al. \cite{glowinski1998ComputMethApplMechEngrg}, who used a variational principle framework for discretization of the NS equations by the finite-element method. The power of the coupled Lagrange multiplier scheme is that it can be straightforwardly adapted to various applications in fluid mechanics. In fact, the approach has been successfully utilized by a number of researchers, namely, by Taira and Colonius \cite{taira2009AIAAJ} for investigation of steady blowing into separated flows behind low-aspect-ratio rectangular wings; by Samanta et al. \cite{samanta2010aiaa} for prediction of the natural convection heat transfer and buoyancy for a hot air balloon; by Yiantsios \cite{yiantsios2012} for the simulation of rigid-particle-laden flows; by Choi et al. \cite{choi2015jfm} for investigation of the forces and unsteady flow structures associated with harmonic oscillations of an airfoil; and recently by Wang and Eldridge \cite{wang2015JCP} for simulating the dynamic interactions between incompressible viscous flows and rigid-body systems.

The present paper reports on our ongoing effort aimed at developing a novel fully pressure-velocity coupled IB solver based on the Lagrange multiplier approach. Similarly to the method presented by Taira and Colonius \cite{taira2007JCP}, the unknown volumetric forces acting at the Lagrangian points are treated as Lagrange multipliers, implicitly coupled with the flow field. The main novelty, however, is that the coupling is implemented on the basis of a fully pressure-velocity coupled direct solver (FPCD) \cite{feldman2009computstruct}, rather than on the projection approach. Therefore, the present method does not require an extra projection-correction step, which, first, significantly boosts the computational efficiency of the time integration process, and, second, allows us to formulate a full Jacobian operator to compute the steady-state solution and then to conduct a linear stability analysis by a shift-invert Arnoldi iteration. To the best of our knowledge, to date the only available approach embedding IB functionality into a linear stability analysis is that due to Giannetti and Luchini \cite{giannettiFM2007}, who utilized an adjoint NS operator (in addition to the direct one) to couple between the immersed body and the surrounding isothermal flow. The present approach does not involve an adjoint operator and is therefore at least twice as efficient in terms of both memory and CPU time consumptions.

Although the developed methodology can formally be applied to both $2D$ and $3D$ flows, only $2D$ configurations were elaborated in the framework of the present study. This is because the algorithm utilizing the direct solver\footnote{We presently use the open source MUMPS solver, http://mumps.enseeiht.fr/.} for $LU$ factorization of the Stokes operator with all no-slip boundaries loses its computational efficiency (due to high memory and time demands) for grids with more than $60$ divisions in each direction \cite{feldman2009computstruct}. The resulting low grid resolution is insufficient for obtaining quantitatively reliable results \cite{gelfgat2007ijnmf}. For $2D$ flows, we show that the developed method preserves its high efficiency for up to $1400^2$ grids. Nevertheless, in the light of the intensive development of modern efficient direct solvers and the likelihood of a rapid increase of computational power, the developed approach will also become attractive for $3D$ simulations in the future.

The paper is organized as follows. In section \ref{NumerForm}, the numerical formulation of the developed methodology is presented. The section includes an introductory description of the previously developed FPCD solver (section \ref{FPCD}), the concepts of IB formalism, based on the Lagrange multipliers approach (section \ref{IbFormalism}), a detailed description of the time marching solver developed in this work (section \ref{TimeStep}), the steady-state solver (section \ref{IbSteadyState}) and the linear stability solver (section \ref{IbStabilitySolver}). Section \ref{Results} presents a detailed verification of all the developed solvers for incident and natural convection incompressible $2D$ flows. The final section presents a summary and the main conclusions of the study.

\section{The numerical formulation}
\label{NumerForm}
The developed numerical methodology, based on the implicit formulation of the IB method and a fully pressure-velocity coupled approach, incorporates three solvers: a time marching solver for the time integration of the NS equations; a steady-state solver based on the full Newton iteration; and a linear stability solver for calculating the necessary part of the whole spectrum of the flow by utilizing the Arnoldi iteration method. All three solvers are based on the previously developed fully pressure-velocity coupled direct (FPCD) solver \cite{gelfgat2007ijnmf,feldman2009computstruct} briefly described here for the sake of completeness.
\subsection{The FPCD solver}
\label{FPCD}
We consider the 2D NS equations for isothermal incompressible flow:
\begin{subeqnarray}
&\nabla\cdot \textbf{\textit{\textbf{u}}} =0,\\
&\frac{\partial{\textbf{\textit{\textbf{u}}}}}{\partial{\textit{t}}}+(\textbf{\textit{\textbf{u}}}\cdotp\nabla){\textbf{\textit{\textbf{u}}}}=
-\nabla{p}+\frac{1}{Re}\nabla^2 \textbf{\textit{\textbf{u}}},
  \label{govern1}
\end{subeqnarray}
where $\textbf{\textit{u}}(u,v)$, $p$, and $Re$ are the non-dimensionalized velocity vector, the pressure field, and the Reynolds number, respectively. By applying a second-order backward finite difference scheme for time discretization, Eqs. (\ref{govern1}) can be rewritten as:
\begin{subeqnarray}
&\nabla\cdot \textbf{\textit{\textbf{$\textbf{u}^{n+1}$}}} =0,\\
&[\frac{1}{Re}\nabla^2 \textbf{\textit{$\textbf{u}$}}-\frac{3}{2\Delta t}\textbf{\textit{$\textbf{u}$}}]^{n+1}-\nabla p = {[(\textbf{\textit{u}}\cdot\nabla)\textbf{\textit{u}}-\frac{2}{\Delta t}\textit{\textbf{u}}]}^n+\frac{1}{2 \Delta t }\textit{\textbf{u}}^{n-1}.
  \label{govern2}
\end{subeqnarray}
Note that all the non-linear terms are taken from the previous time step and moved to the right hand side (RHS) of Eqs. (\ref{govern2}). The system of vector Eqs. (\ref{govern2}) can be compactly rewritten in a block-matrix form as:
\begin{equation}
    \begin{bmatrix}
    H_u & 0 & -\nabla_p^x  \\
    0 & H_v & -\nabla_p^y  \\
    \nabla_u^x &\nabla_v^y & 0
   \end{bmatrix}
   \begin{bmatrix}
    u^{n+1} \\
    v^{n+1} \\
    p
  \end{bmatrix}
  =
   \begin{bmatrix}
    RHS_u^{n-1,n} \\
    RHS_v^{n-1,n} \\
    0
  \end{bmatrix},
  \label{Stokes1}
\end{equation}
 where $\nabla^x$ and $\nabla^y$ are the first derivatives with respect to the $x$ and $y$ coordinates, respectively, $H=\frac{1}{Re}\Delta-3$I$/2\Delta t$ are the corresponding Helmholtz operators acting on $u$ and $v$ velocity components, I is the identity operator, and $\Delta$ is the Laplacian operator. The lower indices correspond to the scalar fields on which an operator acts. The left hand side (LHS) of  Eqs. (\ref{Stokes1}), known as the Stokes operator, is further discetized with a standard staggered mesh second-order conservative finite-volume formulation \cite{patankar1980}. Non-linear terms, moved to the RHS of Eqs. (\ref{Stokes1}), are approximated by the conservative central differencing scheme to exclude the appearance of artificial viscosity (see Ref. \cite{gelfgat2007ijnmf} for the discretization details). Following Refs. \cite{gelfgat2007ijnmf,feldman2009computstruct}, the fully pressure-velocity coupled solution of Eqs. (\ref{Stokes1}) can be obtained by $LU$ factorization of the Stokes operator with a set of suitable boundary conditions for all the velocity components and a single Dirichlet reference point for the pressure field. The discrete Stokes operator remains unchanged during the solution, reducing the  time integration of the NS equations to two backward substitutions at each time step. The high efficiency of the above approach (see Ref. \cite{gelfgat2007ijnmf} for the characteristic computational times) is achieved by utilizing a modern multifrontal direct solver for sparse matrices (MUMPS), exploiting the sparseness of the discrete Stokes operator at both $LU$ factorization and back substitution stages. The FPCD approach formulated in Eqs. (\ref{Stokes1}) can be straightforwardly adjusted to the simulation of natural convection flows, with buoyancy effects being introduced by the Boussinesq approximation and governed by:
\begin{subeqnarray}
&\nabla\cdot \textbf{\textit{\textbf{u}}} =0,\\
&\frac{\partial{\textbf{\textit{\textbf{u}}}}}{\partial{\textit{t}}}+(\textbf{\textit{\textbf{u}}}\cdotp\nabla){\textbf{\textit{\textbf{u}}}}=
-\nabla{p}+Gr^{-0.5}\nabla^2 \textbf{\textit{\textbf{u}}}+ \theta \textit{$\overrightarrow{e_z}$}, \\
&\frac{\partial{\theta}}{\partial{\textit{t}}}+(\textbf{\textit{\textbf{u}}}\cdotp\nabla){\theta}=
Pr^{-1}Gr^{-0.5}\nabla^2\theta,
\label{govern3_1}
\end{subeqnarray}
where $\textbf{\textit{u}}$, $\textit{$\theta$}$, and $p$ correspond to the non-dimensionalized velocity, the temperature and the pressure fields respectively, $Gr$ is the Grashof number, $Pr$ is the Prandtl number, and $\overrightarrow{e_z}$ is the unit vector in the opposite direction to gravity. Discretizing the time by a second-order backward finite difference scheme leads to:
\begin{subeqnarray}
&\nabla\cdot \textbf{\textit{\textbf{$\textbf{u}^{n+1}$}}} =0,\\
&[Gr^{-0.5}\nabla^2 \textbf{\textit{$\textbf{u}$}}-\frac{3}{2\Delta t}\textbf{\textit{$\textbf{u}$}}+\theta\overrightarrow{e_z}]^{n+1}-\nabla p= {[(\textbf{\textit{u}}\cdot\nabla)\textbf{\textit{u}}-\frac{2}{\Delta t}\textit{\textbf{u}}]}^n+\frac{1}{2 \Delta t }\textit{\textbf{u}}^{n-1},\\
&[Pr^{-1}Gr^{-0.5}\nabla^2 \textbf{\textit{$\textbf{$\theta$}$}}-\frac{3}{2\Delta t}\theta]^{n+1}= {[(\textbf{\textit{u}}\cdot\nabla)\textbf{\textit{$\theta$}}-\frac{2}{\Delta t}\textit{\textbf{$\theta$}}]}^n+\frac{1}{2 \Delta t }\textit{\textbf{$\theta$}}^{n-1},
\label{govern3}
\end{subeqnarray}
Then, using the same notations as for Eqs. (\ref{Stokes1}), the compact block-matrix form of the vector Eqs. (\ref{govern3}) reads:
\begin{equation}
    \begin{bmatrix}
    H_u & 0 & 0 & -\nabla_p^x  \\
    0 & H_v & 0 &-\nabla_p^y  \\
    0 & 0 & H_\theta &0  \\
    \nabla_u^x &\nabla_v^y  & 0&0
   \end{bmatrix}
   \begin{bmatrix}
    u^{n+1} \\
    v^{n+1} \\
   \theta^{n+1} \\
    p

  \end{bmatrix}
  =
   \begin{bmatrix}
    RHS_u^{n-1,n} \\
    RHS_v^{n-1,n} \\
    RHS_\theta^{n-1,n} \\
    0
  \end{bmatrix},
  \label{Stokes2}
\end{equation}
where $H_u$=$H_v$=$Gr^{-0.5}\Delta-3$I$/2\Delta t$ are the Helmholtz operators for the scalar momentum equations, and $H_\theta$= $Pr^{-1}Gr^{-0.5}\Delta-3$I$/2\Delta t$ is the Helmholtz operator for the energy equation. All the other notations and the spatial discretization are the same as in Eqs. (\ref{Stokes1}). The discrete differential operators in the LHS of Eqs. (\ref{Stokes1}) and (\ref{Stokes2}) can contain different boundary conditions, and therefore for the general case $H_u\neq H_v $, and $\nabla_u^x \neq \nabla_p^x$, $\nabla_v^y \neq \nabla_p^y$.

\subsection{The immersed boundary formalism}
\label{IbFormalism}
The IB method can be viewed as a "philosophy" for enforcing boundary conditions on the surface of an immersed body of an arbitrary shape. The boundary of an immersed body is typically preset by a series of Lagrangian points $\textbf{\textit{X}}^k$, whose location does not necessarily coincide with the underlying Eulerian grid. Each Lagrangian point is associated with the corresponding discrete volume
$\Delta V^{k}$, such that an ensemble of these volumes forms a thin shell (see Fig. \ref{fig:StagGr}).
\begin{figure}[p]
\centering
\includegraphics[width=0.5\textwidth]{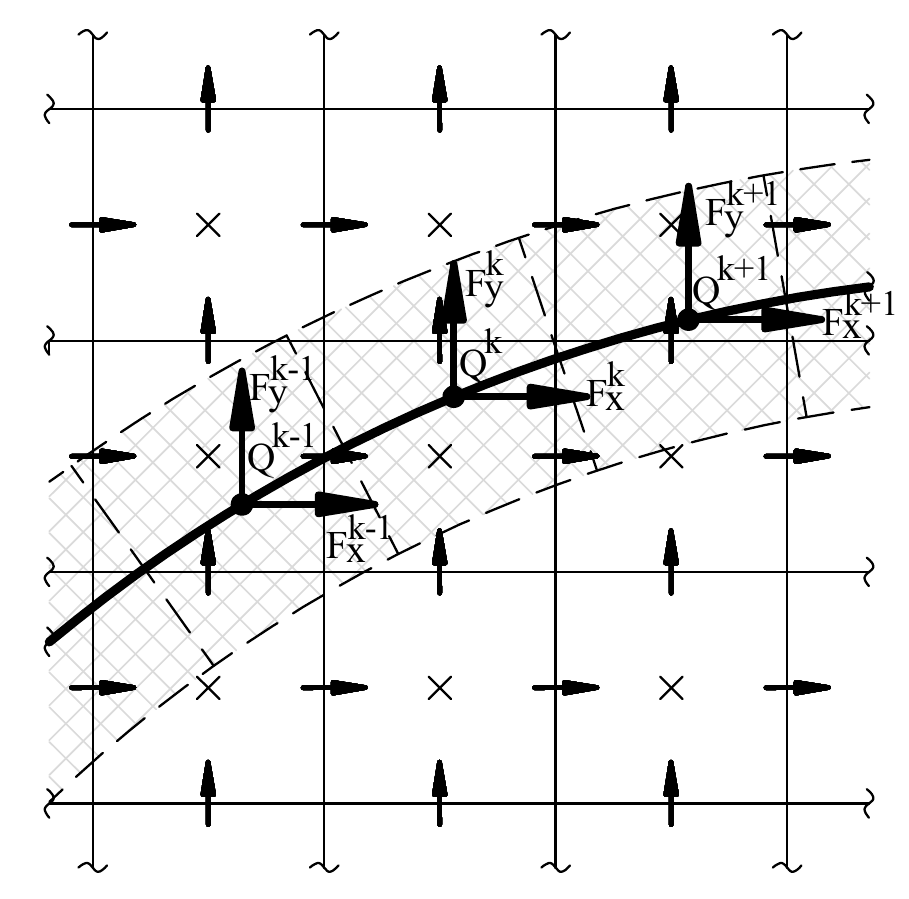}
\caption{Schematic representation of a staggered grid discretization of a two-dimensional computational domain $D$ with a segment of an immersed boundary of a body $B$. A virtual shell, whose thickness is equal to the grid cell width, is  shaded. The horizontal and vertical arrows ($\rightarrow$,$\uparrow$) represent the discrete $u_{i}$ and $v_{i}$  velocity locations, respectively. Pressure $p_{j}$ and temperature $T_{j}$ are applied at the center of each cell $(\times)$. Lagrangian points {\boldmath$\xi$}$^k$($\xi^k$, $\eta^k$)  along $\partial B$ are shown as black circles \textbullet\space where volumetric boundary forces {\boldmath$F$}$^k$=($F^k_x$, $F^k_y$) ($\rightarrow$, $\uparrow$) and volumetric boundary heat sources $Q^{k}$ are applied.
}
\label{fig:StagGr}
\end{figure}
The boundary conditions are enforced by introducing additional functions in the form of volumetric forces, $\textbf{\textit{F}}^{k}$, and heat sources, {$\textit{Q}^{k}$, each associated with the corresponding volume $\Delta V^{k}$. The values of the above functions are not known a priori and are an inherent part of the overall solution in the present implicit formulation. To exchange information between the Eulerian grid and the Lagrangian points, regularization $R$ and interpolation $I$ operators are defined:
\begin{subeqnarray}
&R(\textbf{\textit{F}}^{\textit{k}}(\textit{\textbf{X}}^{\textit{k}}),\textit{Q}^{\textit{k}}(\textbf{\textit{X}}^{\textit{k}}))=\int_{S}(\textbf{\textit{F}}^{\textit{k}}(\textit{\textbf{X}}^{k}),\textit{Q}^{k}(\textbf{\textit{X}}^{k}))
\cdot\delta(\textit{\textbf{x}}_\textit{i}-\textit{\textbf{X}}^{k})dV_S^k,\\
&I(\textbf{\textit{u}}(\textit{\textbf{x}}_\textit{i}),\textit{$\theta$}(\textit{\textbf{x}}_\textit{i}))=\int_{\Omega}(\textbf{\textit{u}}(\textbf{\textit{x}}_i),\theta(\textit{\textbf{x}}_i))\cdot\delta(\textbf{\textit{X}}^{k}-\textit{\textbf{x}}_\textit{i})
dV_{\Omega i},\quad
\label{RegInt}
\end{subeqnarray}
where $S$ corresponds to all the cells belonging to the immersed body surface, $\Omega$ corresponds to a group of flow domain cells located in the close vicinity of the immersed body surface, $d{V}_S^k$ corresponds to the virtual volume surrounding each Lagrangian point $k$, and $dV_{\Omega i}$ is the volume of the corresponding cell of the Eulerian flow domain, whose velocity and temperature values are explicitly involved in enforcing the boundary conditions at point $k$ of the immersed body. The purpose of the regularization operator $R$ is to smear the volumetric forces, $\textbf{\textit{F}}^k$, and heat sources, {$\textit{Q}^k$, on the nearby computational domain by embedding them as sources into the corresponding momentum/energy equations. At the same time, the interpolation operator $I$ imposes no-slip/thermal boundary conditions on the Lagrangian points of the body surface by adding the equations necessary to achieve closure of the overall system. Both operators use convolutions with the Dirac delta function $\delta$ to facilitate an exchange of information between the Lagrangian points of the body surface and the Eulerian grid. The discrete delta function introduced by‎ Roma et al. \cite{roma1999JCP} was used in the present study.
\begin{equation}
  d(r)=\begin{cases}
    \frac{1}{6\Delta r} \Bigg [ 5-3\frac{|r|}{\Delta r}-\sqrt{-3 \Big(1-\frac{|r|}{\Delta r}\Big)^2+1} \Bigg ]& \text{for $0.5\Delta r\leq |r|\leq 1.5\Delta r$},\\
    \frac{1}{3\Delta r} \Bigg [ 1+\sqrt{-3 \Big(\frac{|r|}{\Delta r}\Big)^2+1} \Bigg ]& \text{for $|r|\leq 0.5\Delta r$},\\
    0 & \text{otherwise},\\
  \end{cases}
\end{equation}
where $\Delta r$ is the cell width in the $r$ direction. The chosen delta function was specifically derived for use on staggered grids, and it has been successfully utilized in a number of previous studies \cite{uhlmann2005JCP,taira2007JCP,kempe2012JCP}. The delta function involves only three cells in each computational direction, which is an advantage for computational efficiency. To provide high accuracy, the method utilizes a uniform grid in the vicinity of the immersed body surface. In this region, the distance between the neighboring points of the immersed body surface $\Delta l$ and the width of a grid cell should be approximately the same (i.e., $\Delta l \approx \Delta x = \Delta y$ and $dV_S^k\approx dV_{\Omega i}$). Away from the body, non-uniform discretization can be used. The general discrete forms of the regularization and interpolation operators for 2D geometry are governed by Eqs. (\ref{eq:DiscrRegInterp}):
\begin{subeqnarray}
&(\textit{\textbf{f}}_{i},\textit{q}_{i})=\Delta x^2\sum_{k}(\textit{\textbf{F}}^k,\textit{Q}^k )d(\epsilon^k-x_{i})d(\eta^k-y_{i}),\\
&(\textit{\textbf{U}}^k,\Theta^k)=\Delta x^2\sum_{i}(\textit{\textbf{u}}_i,\textit{$\theta$}_i )d(x_{i}-\epsilon^k)d(y_{i}-\eta^k),
\label{eq:DiscrRegInterp}
\end{subeqnarray}
where $\textit{\textbf{f}}_{i}$,$\textit{q}_{i}$ are the discrete volumetric force and heat source, respectively, defined on a staggered grid ($x_i$, $y_i$) and $\textit{\textbf{U}}^k$,$\Theta^k$ are the
discrete boundary  velocity and temperature, respectively, defined at the $k$-\textit{th} Lagrangian point ($\epsilon^k$, $\eta^k$). Following Peskin \cite{peskin1972JCP} and Beyer and LeVeque \cite{beyer1992}, we used the same delta functions for interpolation and regularization operators.

\subsection{ Implicit immersed boundary FPCD time stepper}
\label{TimeStep}
The discrete pressure $p$ appearing in Eqs. (\ref{govern2} -- \ref{Stokes2}) does not actively participate in time propagation and therefore can be viewed as the Lagrange multiplier that constrains the solenoidal velocity field (see e.g. \cite{grehso1987,taira2007JCP}). It is therefore reasonable to augment the existing Stokes operators (see Eqs. (\ref{Stokes1}) and (\ref{Stokes2})) with the IB functionality by adding an additional set of Lagrange multipliers to enforce the appropriate boundary conditions at the Lagrangian points. Formally, the extended block-matrix form of the Stokes operator for 2D isothermal incompressible flow (see Eqs. (\ref{Stokes1})) is formulated as:
\begin{equation}
\left[
    \begin{array}{c c c;{2pt/2pt} c c }
    H_u &  0 & -\nabla_p^x & R_{F_x} &0 \\
    0 & H_v &  -\nabla_p^y & 0& R_{F_y} \\
    \nabla_u^x &\nabla_v^y &0  &0 &0\\ \hdashline[2pt/2pt]
    I_{u} & 0 & 0 & 0 & 0 \\
    0 & I_{v} & 0 & 0 & 0
   \end{array}
\right]
   \begin{bmatrix}
    u^{n+1} \\
    v^{n+1} \\
    p \\\hdashline[2pt/2pt]
    F_x \\
    F_y
  \end{bmatrix}
  =
   \begin{bmatrix}
    RHS_u^{n-1,n} \\
    RHS_v^{n-1,n} \\
    0 \\\hdashline[2pt/2pt]
    U_b \\
    V_b
  \end{bmatrix}.
  \label{StokesIB1}
  \end{equation}
Here, the vertical and horizontal dashed lines separate between the "original" Stokes operator, located at the top left corner of the matrix, and the additional entries related to the embedded immersed boundary functionality. These additional entries can be formally divided into two types. The first type corresponds to the "weights" of the unknown non-dimensional volumetric forces, $F_x$ and $F_y$,
obtained by applying the regularization operator $R$, smearing the forces over the vicinity of the Lagrangian points. The second type corresponds to the "weights" of the Eulerian velocity components. To precisely impose no-slip boundary conditions, the sum of the above "weights," each multiplied by its Eulerian velocity component, should be equal to the velocities $U_b$ and $V_b$ of the corresponding Lagrangian points. In other words, entries of the second type are nothing more than the additional equations necessary to achieve  closure of the whole system of Eqs. (\ref{StokesIB1}), after the unknowns $F_x$ and $F_y$ have been added. It should be noted that as a result of the utilization of the same Dirac delta functions in both the interpolation $I$ and regularization $R$ operators and the uniform staggered grid in the near vicinity of the immersed body surface, the interpolation and regularization operators are transposed to each other, $R_\textbf{\textit{F}} = I^T_\textbf{\textit{u}}$ . Note also that for all rigid stationary immersed bodies the values of  $U_b$ and $V_b$ are all equal to zero and the extended Stokes operator in Eqs. (\ref{StokesIB1}) does not vary in time. As a result, $LU$ factorization of the the extended Stokes operator should be performed only once at the beginning of the computational procedure. For  moving/deforrming bodies, the location of the Lagrangian points is updated at each time step, requiring modification of the extended Stokes operator (see Eqs. (\ref{StokesIB1})) with its subsequent $LU$ factorization. The factorization can be efficiently performed on a massively parallel machine, taking advantage of the high scalability parallelization built-in into the MUMPS solver \cite{xiaoye2003ACM}.

Using the same notations as for Eqs. (\ref{Stokes2}) and (\ref{StokesIB1}), an extended immersed boundary formulation for the natural convection flow can be written as:
\begin{equation}
\left[
    \begin{array}{c c c c;{2pt/2pt} c c c }
    H_u &  0 & 0 &-\nabla_p^x & R_{F_x} &0 &0 \\
    0 & H_v &  0 &-\nabla_p^y & 0& R_{F_y} &0 \\
    0 & 0&    H_\theta & 0& 0 & 0& R_{Q}  \\
    \nabla_u^x &\nabla_v^y &0 & 0  &0 &0 &0\\ \hdashline[2pt/2pt]
    I_{u} & 0 & 0 & 0 & 0 & 0  &0\\
    0 & I_{v} & 0 & 0 & 0 & 0  &0\\
    0 & 0 & I_\theta  & 0 & 0 & 0 & 0  \\
   \end{array}
\right]
   \begin{bmatrix}
    u^{n+1} \\
    v^{n+1} \\
    \theta^{n+1} \\
    p \\ \hdashline[2pt/2pt]
    F_x \\
    F_y \\
    Q
  \end{bmatrix}
  =
   \begin{bmatrix}
    RHS_u^{n-1,n} \\
    RHS_v^{n-1,n} \\
    RHS_\theta^{n-1,n} \\
    0 \\\hdashline[2pt/2pt]
    U_b \\
    V_b \\
    \Theta
  \end{bmatrix}.
  \label{StokesIB2}
  \end{equation}

Similarly to the Eqs. (\ref{StokesIB1}), the "original" Stokes operator located at the top left corner of the block-matrix form is separated by the vertical and horizontal dashed lines from the immersed boundary entries. The $R_Q$ entries correspond to the "weights"  of the unknown non-dimensional volumetric heat sources smeared over the vicinity of the corresponding Lagrangian points by the regularization operator $R$, whereas the $I_\theta$ entries are the "weights" of the Eulerian temperatures, imposing Dirichlet boundary conditions at the neighboring Lagrangian points.

In most thermal problems, precise estimation of the average $\overline{Nu}$ number is of significant practical importance and is particulary critical for the present implementation of the IB method, which relies on a uniform Cartesian grid. As a result, a further refining of the Eulerian grid adjacent to the immersed boundary for a more precise resolution of the thinnest boundary layers is not practical. An alternative way to obtain an accurate estimation the $Nu$ number is to express the unknown Lagrangian non-dimensional volumetric heat sources in terms of the temperature gradients in the direction normal to the immersed boundary as:
\begin{equation}
 Q=\frac{1}{Pr\sqrt{Gr}\Delta x}\frac{\partial{\theta}}{\partial{\textit{\textbf{n}}}},
\label{QModif}
\end{equation}
where $\Delta x$= $\Delta y$ is the dimension of the uniform Eulerian grid in the vicinity of the immersed surface. Following \cite{ren2012compufluid}, the $\overline{Nu}$ value averaged over the surface of the immersed body reads:
\begin{equation}
\overline{Nu}=\frac{1}{2}\Big ( \sum_{k=1}^{M}\frac{\partial{\theta}}{\partial{\textit{\textbf{n}}}}\triangle x \Big)_k,
\label{NuAv}
\end{equation}
where the local $\frac{\partial{\theta}}{\partial{\textit{\textbf{n}}}}$ values at every point $1\leq k\leq M$ of the immersed body are provided by the solution of Eqs. (\ref{StokesIB2}), reformulated in terms of the temperature gradients in the direction normal to the body surface. Following the same principle, the drag $C_d$ and the lift $C_l$ coefficients can be be obtained by:
\begin{equation}
(C_d,C_l)=-2\sum_{k=1}^{M}(F_{x_{k}},F_{y_{k}})/\rho U_{\infty}d,
\label{NuAv}
\end{equation}
where $F_{x_{k}}$ and $F_{y_{k}}$ are an intrinsic part of the overall solution obtained at every point $k$ of the immersed body and $\rho U_{\infty}d=1$ for the presently used normalization.

The above immersed boundary formulation embedded into the FPCD time stepper can be seen as an extension of the algorithm recently developed by Taira and Colonius \cite{taira2007JCP}, who coupled unknown volumetric forces acting at the Lagrangian points with an intermediate non-solenoidal velocity field, which  must then be further projected to the divergence free subspace by a projection-correction step. Based on the full  pressure-velocity coupling, the present direct method does not require the projection-correction step, which is an advantage for computational efficiency.
\subsection{ Steady-state immersed boundary FPCD solver}
\label{IbSteadyState}
A steady isothermal incompressible flow with an embedded immersed boundary functionality is governed by the following continuity and momentum equations:
\begin{subeqnarray}
&\nabla\cdot \textbf{\textit{\textbf{u}}} =0,\\
&(\textbf{\textit{\textbf{u}}}\cdotp\nabla){\textbf{\textit{\textbf{u}}}}+\nabla{p}-\frac{1}{Re}\nabla^2 \textbf{\textit{\textbf{u}}}-R_\textbf{\textit{F}}=0,\\
&I(\textit{\textbf{u}})-\textbf{\textit{U}}_b=0,
  \label{SteadyIB}
\end{subeqnarray}
where $R_\textbf{\textit{F}}$ and $I(\textit{\textbf{u}})$ are additional entries resulting from applying the regularization $R$ and interpolation $I$ operators. Note that the steady-state formulation formally treats the flow around an immersed body in the same way as its unsteady analog given by Eqs. (\ref{StokesIB1}). All the differential operators of Eqs.(\ref{SteadyIB}) are subsequently discretized in space by the standard staggered grid second-order conservative finite-volume method (in the same way as in the corresponding unsteady formulation). All the additional entries related to the IB formulation are discretized by using discrete Dirac delta functions. The discretized Eqs. (\ref{SteadyIB}) summarized in a compact block-matrix form in Eqs. \ref{NewtonIB1} are then solved by the Newton-Raphson method.
\begin{equation}
\left[
    \begin{array}{c c c;{2pt/2pt} c c }
    J_x &  0 &  J_p & R_{F_x} &0 \\
    0 & J_y &  J_p & 0& R_{F_y} \\
    J_u &J_v &0  &0  &0\\ \hdashline[2pt/2pt]
    I_{u} & 0 & 0 & 0 & 0 \\
    0 & I_{v} & 0 & 0 & 0
   \end{array}
\right]
   \begin{bmatrix}
    \delta(u) \\
    \delta(v) \\
    \delta (p) \\\hdashline[2pt/2pt]
    \delta(F_x) \\
    \delta(F_y)
  \end{bmatrix}
  = -
   \begin{bmatrix}
    Fn_x\enspace ${\raisebox{-.6ex}{${ \Large\rotatebox{90}{\Kutline}}$}}$ - $$\sum_{k}^{} R_{kF_x} $$  \\
    Fn_y\enspace${\raisebox{-.6ex}{${  \Large\rotatebox{90}{\Kutline}}$}}$ - $$\sum_{k}^{} R_{kF_y} $$  \\
    Fn_p \\\hdashline[2pt/2pt]
    $$\sum_{i}^{} I_{iu_x} $$ -U_{b_x}\\
    $$\sum_{i}^{} I_{iu_y}$$  -U_{b_y}
  \end{bmatrix},
  \label{NewtonIB1}
  \end{equation}
where $J_x$, $J_y$, $J_p$, $J_u$, $J_v$ entries of Jacobian $\textit{\textbf{J}}$  correspond to the discrete linearized terms of the "original" (without IB functionality) momentum and continuity equations, with the corresponding discrete right-hand sides $Fn_x$, $Fn_y$, $Fn_p$ being calculated at the iteration $n$. The additional entries $R_{\textbf{\textit{F}}}$ and $I_\textbf{\textit{u}}$ of  the Jacobian operator, related to the embedded IB formulation, are separated by the horizontal and vertical dashed lines. The IB entries also contribute to the RHS of Eqs. (\ref{NewtonIB1}). The sums of smeared volumetric forces $\textbf{\textit{F}}_k$ and interpolated velocities $\textbf{\textit{u}}_i$, both calculated at iteration $n$, are added to the corresponding right hand sides of the momentum equations and to the complementary interpolation relations. Here, the indexes $i$ and $k$ represent the total number of Eulerian and Lagrangian points, respectively, participating in the summation.

The developed steady-state IB solver can be straightforwardly adjusted to the steady-state solution of the natural convection flow, governed by:
\begin{subeqnarray}
&\nabla\cdot \textbf{\textit{\textbf{u}}} =0,\\
&(\textbf{\textit{\textbf{u}}}\cdotp\nabla){\textbf{\textit{\textbf{u}}}}
+\nabla{p}-Gr^{-0.5}\nabla^2 \textbf{\textit{\textbf{u}}}- \theta \textit{$\overrightarrow{e_z}$}-R_\textbf{\textit{F}}=0, \\
&(\textbf{\textit{\textbf{u}}}\cdotp\nabla){\theta}-Pr^{-1}Gr^{-0.5}\nabla^2\theta-R_\textit{Q}=0, \\
&I(\textit{\textbf{u}})-\textbf{\textit{U}}_b=0, \\
&I(\textit{$\theta$})-\textit{$\Theta$}_b=0
\label{SteadyIB2}
\end{subeqnarray}
where  the Boussinesq approximation is utilized for simulating the buoyancy effects, and again $R_\textbf{\textit{F}}$, $R_\textbf{\textit{Q}}$, $I(\textit{\textbf{u}})$, $I(\textit{$\theta$})$ are the additional entries stemming from applying the regularization $R$ and interpolation $I$ operators. Utilizing the same spatial discretization and Dirac delta functions as for Eqs. (\ref{NewtonIB1}), the discretized Eqs. (\ref{SteadyIB2}) are solved by the Newton-Raphson method, whose compact block-matrix form reads:
\begin{equation}
\left[
    \begin{array}{c c c c;{2pt/2pt} c c c }
    J_x &  0 & 0&  J_p & R_{F_x}  &0 &0 \\
    0 & J_y & 0& J_p & 0& R_{F_y} &0 \\
    0 & 0 & J_\theta & 0 &0 &0 &R_{Q} \\
    J_u &J_v &0  &0 &0 &0 &0\\ \hdashline[2pt/2pt]
    I_{u} & 0 & 0 & 0 &0 & 0 &0\\
    0 & I_{v} & 0 & 0 &0 & 0 &0\\
    0 & 0 &  I_{\theta}  & 0 &0 & 0 &0
   \end{array}
\right]
   \begin{bmatrix}
    \delta(u) \\
    \delta(v) \\
    \delta(\theta) \\
    \delta (p) \\\hdashline[2pt/2pt]
    \delta(F_x) \\
    \delta(F_y)\\
    \delta(Q)\\
  \end{bmatrix}
  = -
   \begin{bmatrix}
    Fn_x\enspace ${\raisebox{-.6ex}{${ \Large\rotatebox{90}{\Kutline}}$}}$ - $$\sum_{k}^{} R_{kF_x} $$  \\
    Fn_y\enspace${\raisebox{-.6ex}{${  \Large\rotatebox{90}{\Kutline}}$}}$ - $$\sum_{k}^{} R_{kF_y} $$  \\
    Fn_\theta\enspace${\raisebox{-.6ex}{${  \Large\rotatebox{90}{\Kutline}}$}}$ - $$\sum_{k}^{} R_{kQ} $$  \\
    Fn_p \\\hdashline[2pt/2pt]
    $$\sum_{i}^{} I_{iu_{x}} $$ -U_{b_x}\\
    $$\sum_{i}^{} I_{iu_{y}}$$  -U_{b_y}\\
    $$\sum_{i}^{} I_{i\theta}$$  -\Theta_{b}\\
  \end{bmatrix},
  \label{NewtonIB2}
  \end{equation}
The compact block-matrix form of Eqs. (\ref{NewtonIB2}) bears a striking resemblance to that determined by Eqs. (\ref{NewtonIB1}) (corresponding to the isothermal flow), the only exceptions being the additional entries related to the energy equations and to the Diriclet temperature boundary conditions applied to the immersed surface. As was done in the time integration analysis, the volumetric heat sources $Q$ can be expressed in terms of the normal temperature gradients $\frac{\partial{\theta}}{\partial{\textit{\textbf{n}}}}$ (see Eq. (\ref{QModif})) required for the precise estimation of the $\overline{Nu}$ value. Note that all the entries related to the IB functionality (i.e., $I_u$, $I_v$, $I_\theta$, $R_{Fx}$, $R_{Fy}$, $R_{Q}$) are linear and therefore have the same form both in the Stokes operator (see Eqs. (\ref{StokesIB1}) and (\ref{StokesIB2})) and in the corresponding Jacobian operator (see Eqs. (\ref{NewtonIB1}) and (\ref{NewtonIB2})).

\subsection{ Linear stability immersed boundary FPCD solver}
\label{IbStabilitySolver}
For the sake of conciseness, only equations for the linear stability analysis of the natural convection flow will be derived in this section. The equations for the linear stability of the isothermal flow can be obtained by a straight-forward omission of the energy equations and the temperature terms in the corresponding momentum equations. The linear stability eigenproblem is formulated by assuming infinitesimally small perturbations in the form of \{$\widetilde{\textbf{\textit{u}}}$(\textit{x},\textit{y}), $\widetilde{\theta}$(\textit{x},\textit{y}), $\widetilde{p}$(\textit{x},\textit{y}), $\widetilde{\textbf{\textit{F}}}$(\textit{x},\textit{y}), $\widetilde{Q}$(\textit{x},\textit{y})\}$e^{\lambda t}$ around the steady state flow $\textit{\textbf{U}}$, \textit{$\Theta$}, $P$, $\textit{\textbf{F}}$, \textit{$Q$}, as follows:
\begin{subeqnarray}
&\lambda\tilde{\textbf{\textit{u}}}=- (\textbf{\textit{\textbf{U}}}\cdot\nabla)\tilde{\textbf{\textit{u}}}-(\tilde{\textbf{\textit{\textbf{u}}}}\cdot\nabla){\textbf{\textit{U}}}
-\nabla{\tilde{p}}+Gr^{-0.5}\nabla^2 \tilde{\textbf{\textit{\textbf{u}}}}- \tilde{\theta} \textit{$\overrightarrow{e_z}$}+R_{\tilde{\textbf{\textit{F}}}}, \\
&\lambda \tilde{\theta}=-(\textbf{\textit{\textbf{U}}}\cdot\nabla)\tilde{\theta}-(\tilde{\textbf{\textit{\textbf{u}}}}\cdot\nabla)\Theta+Pr^{-1}Gr^{-0.5}\nabla^2\tilde{\theta}+R_{\tilde{\textit{Q}}}, \\
&0=\nabla\cdot \tilde{\textbf{\textit{u}}},\\
&0=I(\tilde{\textit{\textbf{u}}}), \\
&0=I(\tilde{\theta}),
\label{StabilityIB2}
\end{subeqnarray}
or in a block-matrix form as:

\begin{equation}
\lambda\textbf{\textit{B}}
\left[
   \begin{array}{c}
    \tilde{\textbf{\textit{u}}} \\
    \tilde{\theta} \\
    \tilde{p} \\
    \tilde{\textbf{\textit{F}}} \\
    \tilde{\textit{Q}} \\

   \end{array}
\right]
  = \textit{\textbf{J}}
   \left[
   \begin{array}{c}
    \tilde{\textbf{\textit{u}}} \\
   \tilde{\theta} \\
   \tilde{p} \\
    \tilde{\textbf{\textit{F}}} \\
    \tilde{\textit{Q}} \\

   \end{array}
\right],
  \label{StabilityIB2Block}
  \end{equation}
where \textit{\textbf{J}} is the Jacobian matrix calculated from the RHS of Eqs. (\ref{StabilityIB2}), and \textbf{\textit{B}} is the diagonal matrix  whose diagonal elements, corresponding to the values of $\widetilde{\textbf{\textit{u}}}$, $\widetilde{\theta}$ are equal to unity, and whose diagonal elements, corresponding to  $\widetilde{p}$, $\widetilde{\textbf{\textit{F}}}$, $\widetilde{Q}$, are equal to zero. Note that for Cartesian coordinates and the staggered uniform grid in the vicinity of immersed body surface, the discrete Jacobians, \textit{\textbf{J}} of  Eqs. (\ref{NewtonIB2}) and (\ref{StabilityIB2Block}) are the same.
The generalized eigenproblem (\ref{StabilityIB2Block}) cannot be directly transformed into a standard eigenproblem, since $det(\textbf{\textit{B}})=0$; instead it is solved in a shift-invert mode:
\begin{equation}
(\textit{\textbf{J}}-\sigma\textbf{\textit{B}})^{-1}\textbf{\textit{B}}
\left[
   \begin{array}{c}
    \tilde{\textbf{\textit{u}}} \\
    \tilde{\theta} \\
    \tilde{p} \\
    \tilde{\textbf{\textit{F}}} \\
    \tilde{\textit{Q}} \\

   \end{array}
\right]
  = \mu
   \left[
   \begin{array}{c}
    \tilde{\textbf{\textit{u}}} \\
    \tilde{\theta} \\
    \tilde{p} \\
    \tilde{\textbf{\textit{F}}} \\
    \tilde{\textit{Q}} \\
   \end{array}
\right], \qquad
  \mu= \frac{1}{\lambda-\sigma}
  \label{StabilityIB2BlockShift}
  \end{equation}

The solution is based on a standard Arnoldi iteration implemented within an open source ARPACK package \footnote {http://www.caam.rice.edu/software/ARPACK/}, providing the dominant eigenvalue (i.e., the eigenvalue with the largest modulus). In a linear stability analysis, we are typically interested in finding the critical value of the control parameter (e.g., $Gr_{cr}$ or $Re_{cr}$ numbers) at which $Real(\lambda)=0$ (to a prescribed precision), where $\lambda$ is the leading eigenvalue. The dominant eigenvalue $\mu$ can be related to the leading eigenvalue $\lambda$ (i.e., that of a zero real part) when the approach is applied to a shift-invert problem, where $\sigma$ is a complex shift (see Eqs. (\ref{StabilityIB2BlockShift})). To converge, the approach requires that the complex shift $\sigma$\footnote {Typically $\sigma$ is a pure imaginary number, since $Real(\lambda) \rightarrow 0$ at $Gr\approx Gr_{cr}$, or $Re\approx Re_{cr}$}  be close to the $\lambda$ value, whose imaginary part $Im(\lambda)$ corresponds to the critical angular oscillating frequency, $\omega_{cr}$. The value of $\omega_{cr}$ is either known a priori (for benchmark problems) or can be estimated by a series of successive direct numerical simulations of the slightly bifurcated flow.

The present linear stability approach extends the algorithm presented by Gelfgat \cite{gelfgat12007ijnmf}, with an IB functionality. Theoretically, no specific restrictions are imposed either on the number of bodies or on their shape. However, the method requires that the body boundaries do not touch or intersect and that the minimal distance between neighboring bodies is at least the size of a single grid cell. The solution procedure is as follows. First, the steady-state solution is calculated by the Newton method for the given value of the control parameter ($Gr$ or $Re$ numbers). Then, the linear stability analysis is performed by utilizing a shift-invert Arnoldi iteration (see Eqs. (\ref{StabilityIB2BlockShift})). The corresponding egenvalue problem is solved by a secant method, providing a precise value for the critical control parameter. The overall process requires numerous solutions of large systems of linear equations, which should be performed at each step of the Newton method and while building the Krylov basis for the Arnoldi iteration. Typically no more than ten iterations are required for the calculation of the steady-state solution (by the Newton method), while the shift-invert Arnoldi iteration needs $O(10^4)$ iterations to converge, thus comprising the key issue determining the computational efficiency of the whole process.

Next, to efficiently implement the product of the operator $(\textit{\textbf{J}}-\sigma\textbf{\textit{B}})^{-1}\textbf{\textit{B}}$ by the vector $[\tilde{\textbf{\textit{u}}}, \tilde{\theta}, \tilde{p}, \tilde{\textbf{\textit{F}}}, \tilde{\textit{Q}}]^T$ required at each Arnoldi iteration step, we exploit the fact that the operator $(\textit{\textbf{J}}-\sigma\textbf{\textit{B}})^{-1}\textbf{\textit{B}}$ does not change during the building of the Krylov basis for the Arnoldi iteration (see Eqs. (\ref{StabilityIB2BlockShift})). The product implementation is simply a solution $\textit{\textbf{X}}$ of the linear system $(\textit{\textbf{J}}-\sigma\textbf{\textit{B}})\textit{\textbf{X}}=\textbf{\textit{B}}[\tilde{\textbf{\textit{u}}}, \tilde{\theta}, \tilde{p}, \tilde{\textbf{\textit{F}}}, \tilde{\textit{Q}}]^T$. By utilizing the direct solver MUMPS, the $LU$ decomposition of the operator $(\textit{\textbf{J}}-\sigma\textbf{\textit{B}})$ is performed once at the beginning of the process, and then each vector of the Krylov basis is obtained by just two subsequent back substitutions, whose complexity is comparable to that of  matrix-vector multiplication. Note also that the overall performance is  additionally boosted by being a  $(\textit{\textbf{J}}-\sigma\textbf{\textit{B}})$  sparse matrix. The superiority of the above approach over algorithms utilizing modern Krylov-subspace-based iteration methods (e.g.,  preconditioned GMRES and BiCGstab) for building  the Krylov basis for the Arnoldi iteration was extensively discussed in \cite{gelfgat12007ijnmf} for natural convective flows in cavities. In the present study, we successfully extended the approach by embedding the IB functionality and applied it to a linear stability analysis of both open and confined flows.

\section{ Results}
\label{Results}
\subsection{ Unsteady flow: periodic incident flow around two horizontally aligned circular cylinders}
\label{UnstableTandem}
Verification of the developed implicit IB FPCD time stepper was first performed for simulation of the secondary instabilities in the flow around a tandem arrangement of two equal horizontally aligned cylinders of diameter $d$, as shown in Fig. \ref{fig:HorTandemSchematic}.
 \begin{figure}
\centering
\includegraphics[width=0.7\textwidth,clip=]{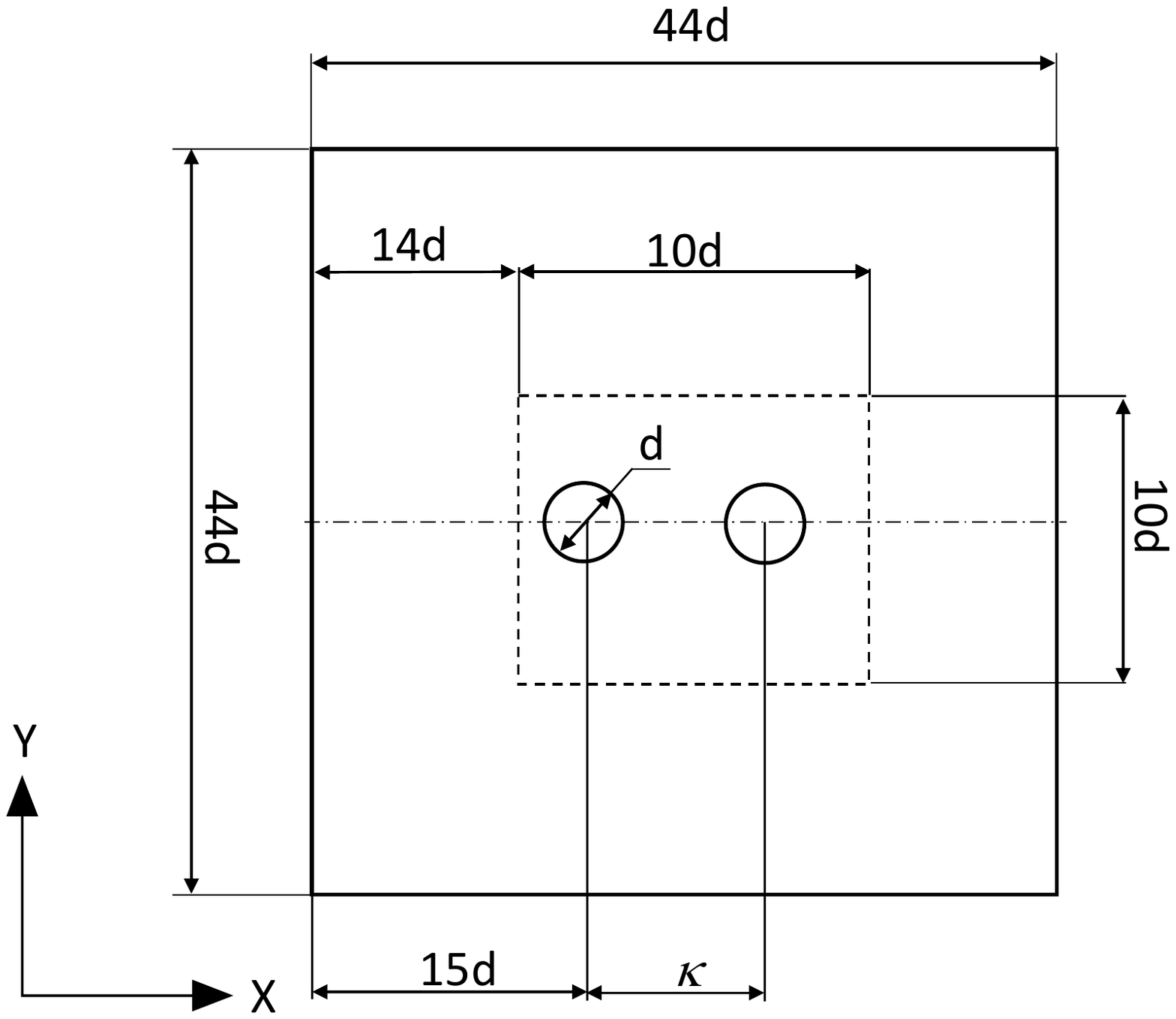}
\caption{Schematic representation of the geometrical model and discretization of the computational domain for the flow around two horizontally aligned cylinders arranged in tandem.}
\label{fig:HorTandemSchematic}
\end{figure}
All the simulations were performed in a square computational domain of size $44$ $d$ in each direction. The two cylinders were centered in the vertical direction,  while a distance equal to $16$ $d$ was set between the center of the forward cylinder and the inlet boundary of the domain. The computational domain was discretized by a non-uniform $1400\times1400$ mesh in the following manner: a square subregion of size $10$ $d$ in each direction, confining the pair as shown in Fig. \ref{fig:HorTandemSchematic}, was discretized by a uniform $1000\times1000$ mesh with a grid step equal to $\Delta x = \Delta y =0.01$. The mesh was built out of the square subregion (see Fig. \ref{fig:HorTandemSchematic}) by gradually increasing $\Delta x$ and $\Delta y$ grid steps, which finally attain the values of  $\Delta x \approx \Delta y \approx 0.22$ at all boundaries of the computational domain. Three configurations, each corresponding to a different center-to-center distance of $\kappa=L_x/d=[1.5,2.3,5]$ between the cylinders, were simulated for the value of $Re=U_{\infty}d/\nu=200$, thus representing three different vortex shedding scenarios, as shown in Fig. \ref{fig:tandemTime}. Here, $\nu$ is the kinematic viscosity of the fluid. The solutions were obtained with the following set of boundary conditions:
\begin{subeqnarray}
    u_x(x=0,y)=u_x(x,y=0, y=44 d)=1, \\
    u_y(x=0,y)=u_y(x,y=0, y=44 d)=0, \\
    p(x=44 d,y)=0,\\
     \frac{\partial \textbf{u} }{\partial t} +\frac{\partial \textbf{u}}{\partial x}(x=44 d,y)=0.
\label{BC}
\end{subeqnarray}
Note that Eq. \ref{BC}d determines the convective boundary condition at the outlet, allowing the  vorticity to exit the domain freely \cite{taira2007JCP}. The obtained results (see Fig. \ref{fig:tandemTime}) are in excellent qualitative and quantitative agreement with the corresponding flow characteristics obtained by Carmo et al. \cite{Carmo2010JFM}, who made a distinction  between the three observed shedding scenarios, classifying them as SG (symmetric in the gap) for $\kappa=1.5$, AG (alternating in the gap) for $\kappa=2.3$, and WG (wake in the gap) for $\kappa=5$. Note the interesting phenomenon of the drag inversion, extensively elaborated in \cite{carmo2010PhysFl} and characterized by a negative to positive change in the value of drag coefficient when the shedding regime changes from AG to WG.

\begin{figure}
\centering
    \subfigure[]
    {
        \includegraphics[width=0.45\textwidth,clip=]{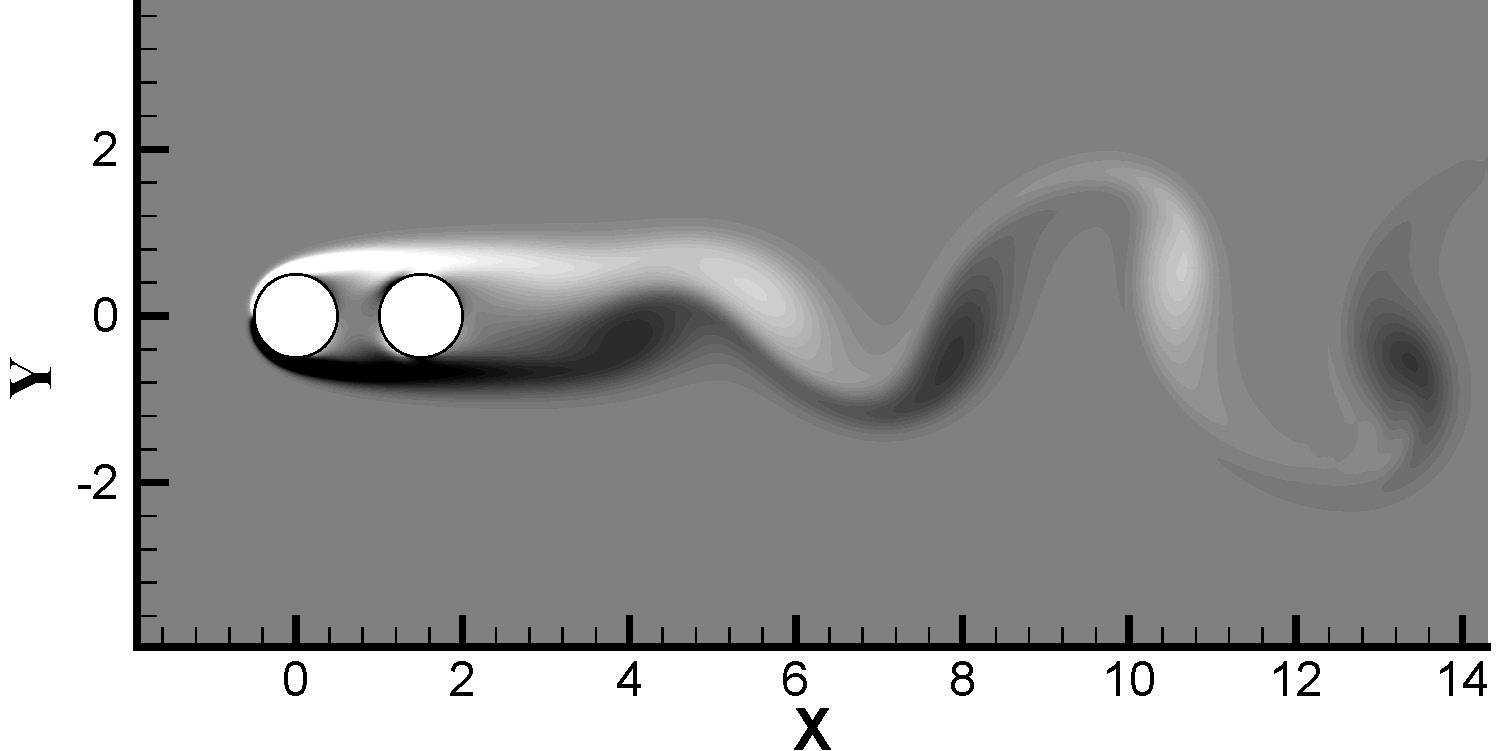}
    }
    \subfigure[]
    {
        \includegraphics[width=0.45\textwidth,clip=]{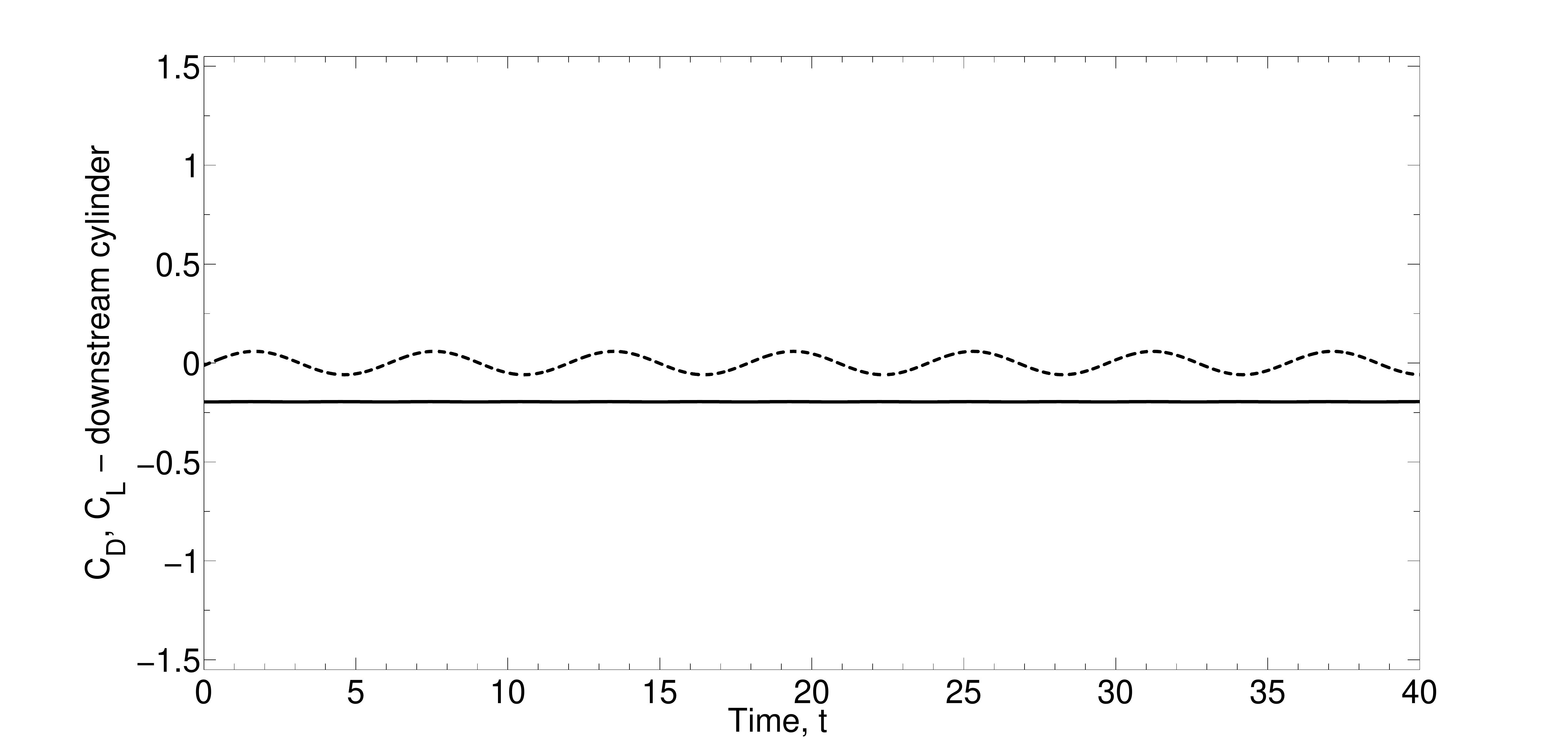}
    }
    \subfigure[]
    {
        \includegraphics[width=0.45\textwidth,clip=]{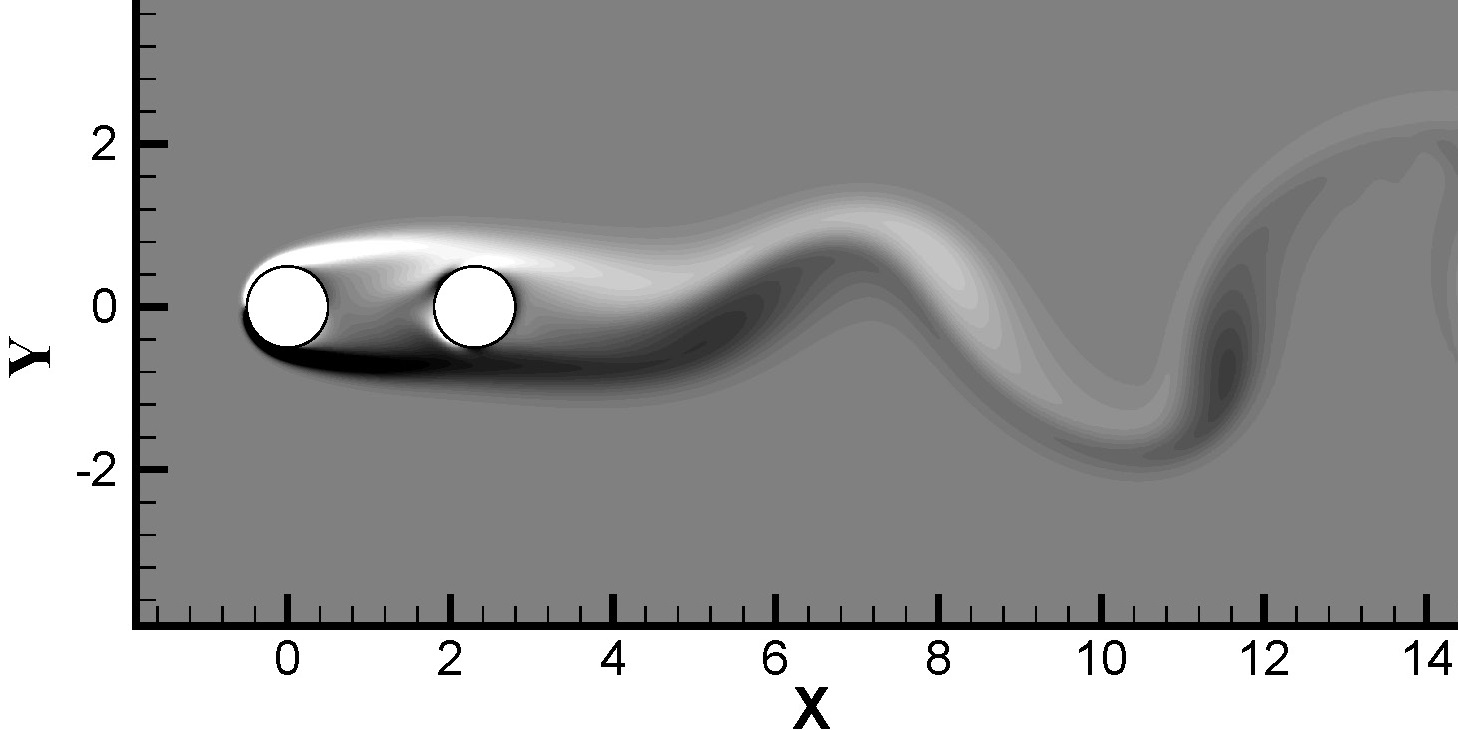}
    }
    \subfigure[]
    {
        \includegraphics[width=0.45\textwidth,clip=]{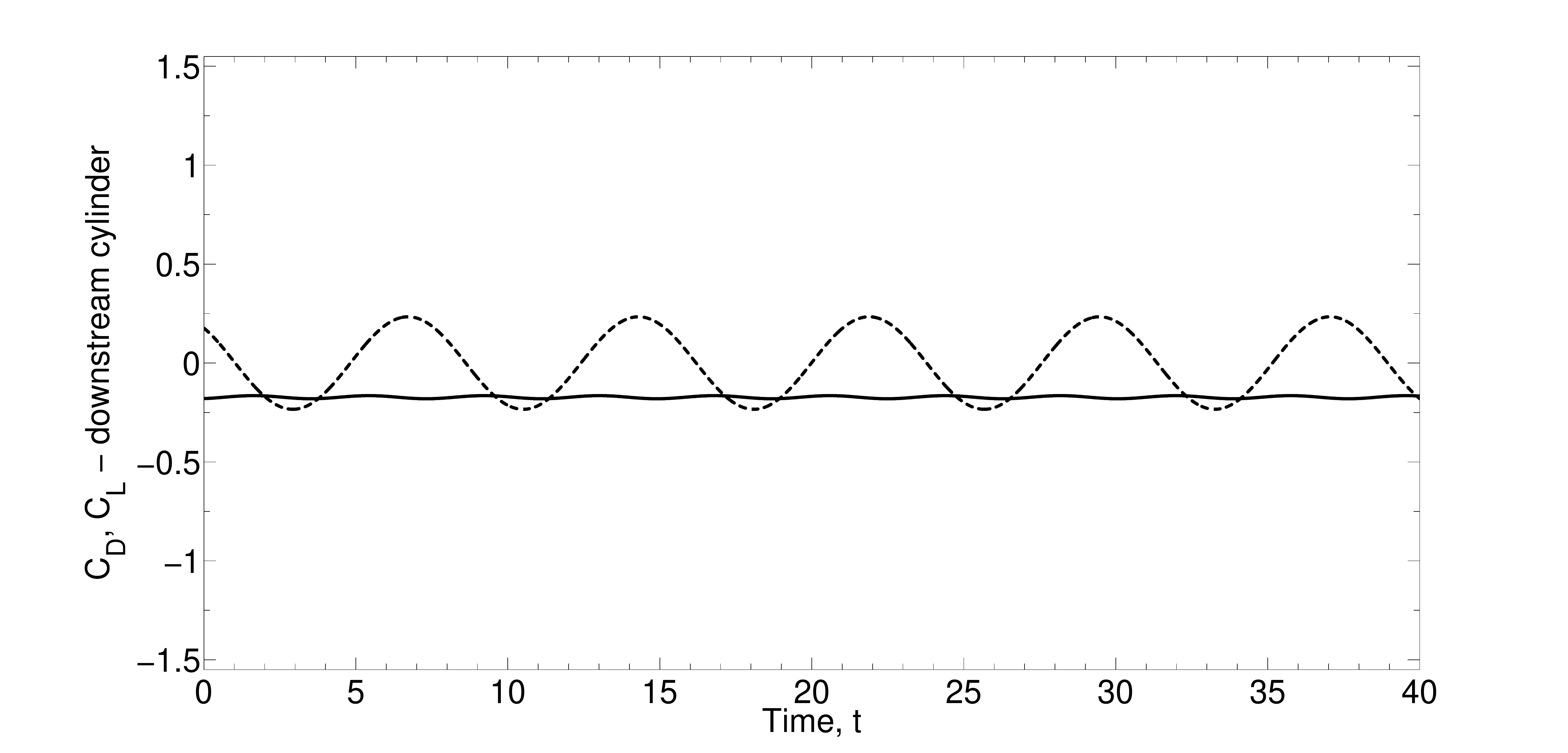}
    }
    \subfigure[]
    {
        \includegraphics[width=0.45\textwidth,clip=]{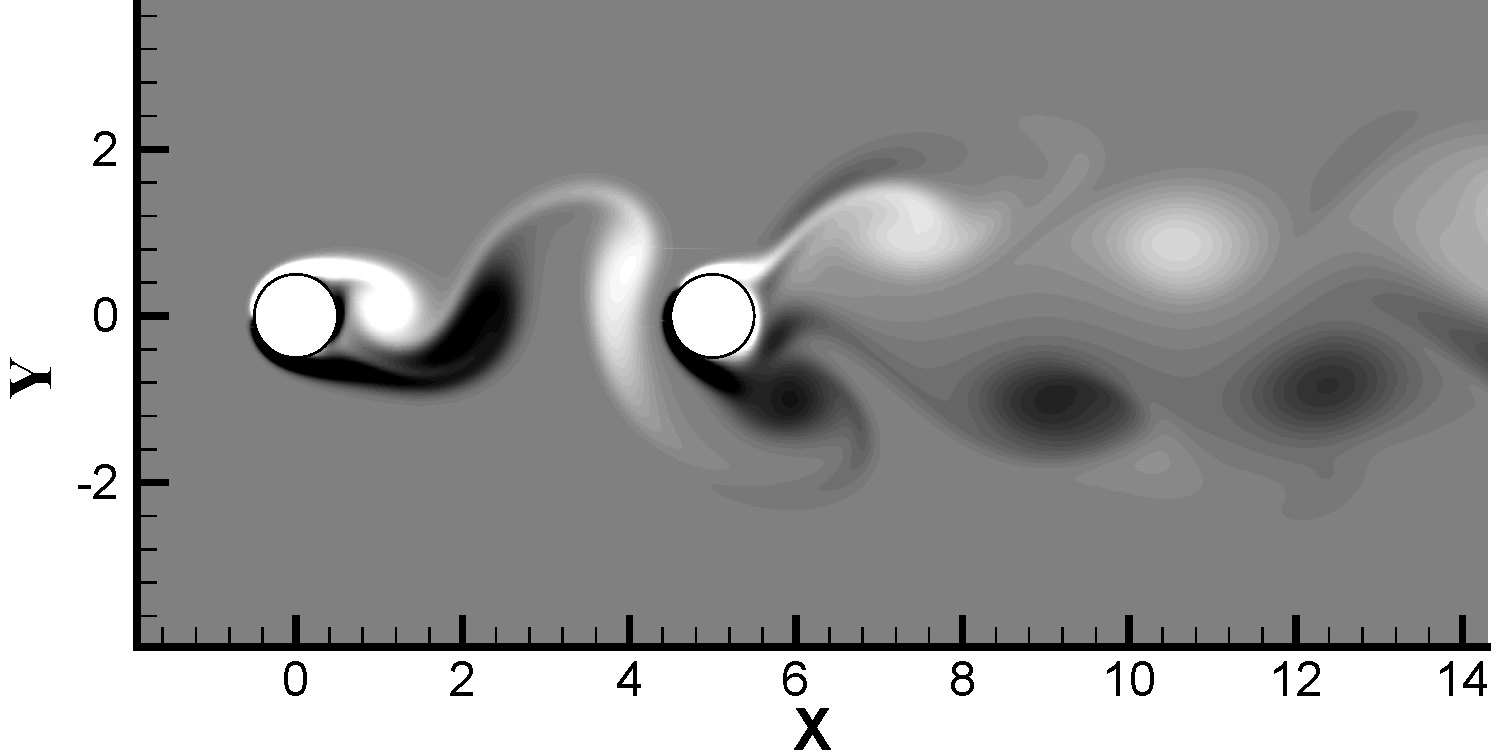}
    }
    \subfigure[]
    {
        \includegraphics[width=0.45\textwidth,clip=]{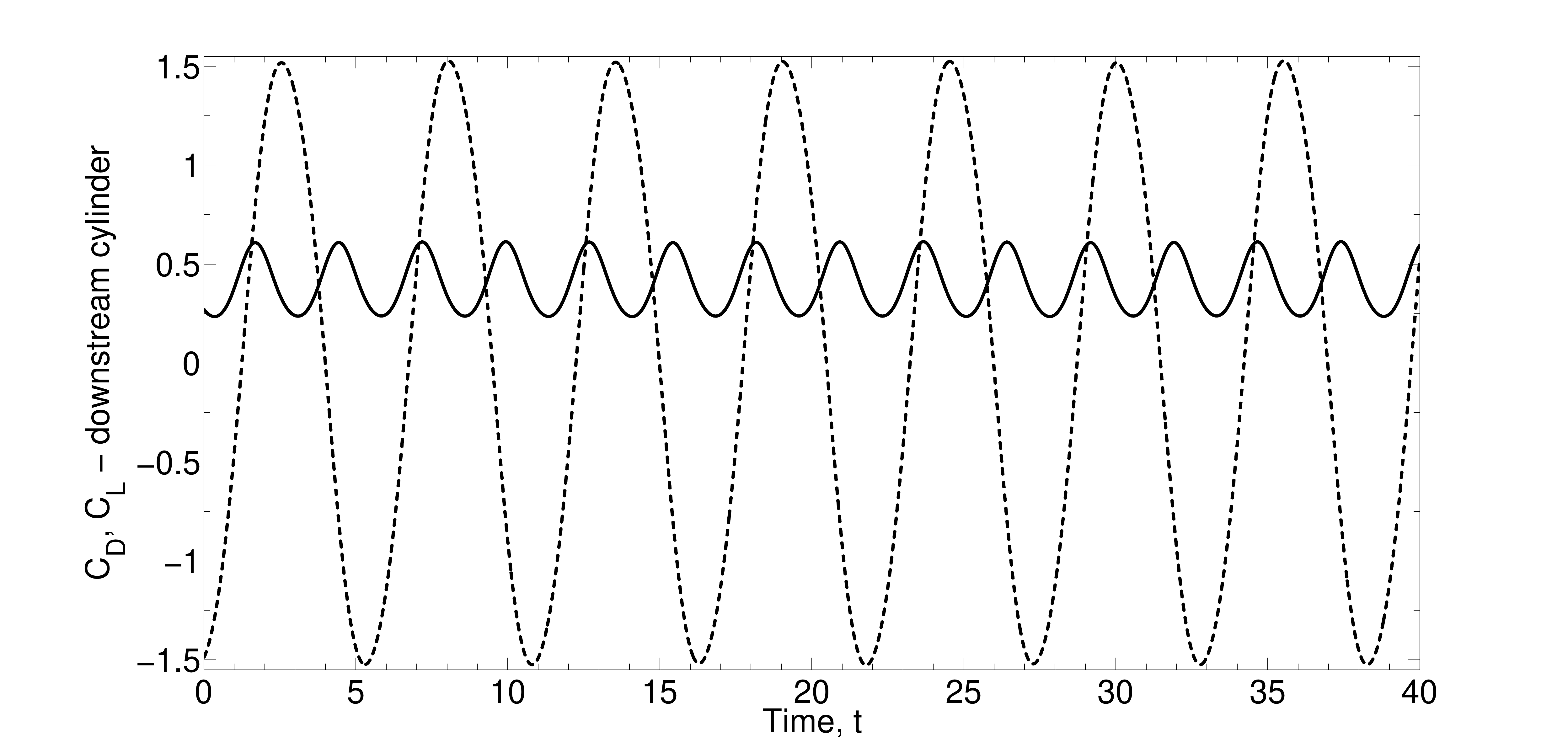}
    }

\caption{Instantaneous vorticity contours, with drag coefficient, $C_D$ (solid line) and lift coefficient $C_L$ (dashed line),  measured on the downstream cylinder at $Re=200$ for: (a) $\kappa=1.5$ -- regime $SG$; (b) $\kappa=2.3$ -- regime $AG$ ; (c) $\kappa=5$ -- regime $WG$.}
\label{fig:tandemTime}
\end{figure}

\subsection{ Unsteady flow: periodic natural convection flow around two vertically aligned circular cylinders}
\label{UnsteadyNatConvect}
The next verification study of the developed time stepper was related to the simulation of unsteady natural convection flow around two cylinders confined by a square cavity (see Fig. \ref{fig:VerTandemSchematic}). The ratio between the cylinder diameter, $d$, and the cavity side length, $L$, is equal to $d/L=0.2$. The cylinders are aligned along the cavity's vertical centerline and are symmetrically distanced  from the cavity's horizontal centerline. The distance $\delta$ between the cylinder centers, normalized by the cavity side length $L$, is equal to $\delta=0.5$. Both cylinders are held at a constant hot temperature $\theta_H=1$, whereas all the cavity boundaries are held at a constant cold temperature, $\theta_C=0$. The force of gravity acts in the $-\hat{y}$ direction.
\begin{figure}
\centering
\caption{Schematic repredentation of a geometrical model of the computational domain for the natural convection flow around vertically aligned cylinders confined by a square cavity.}
\includegraphics[width=0.7\textwidth,clip=]{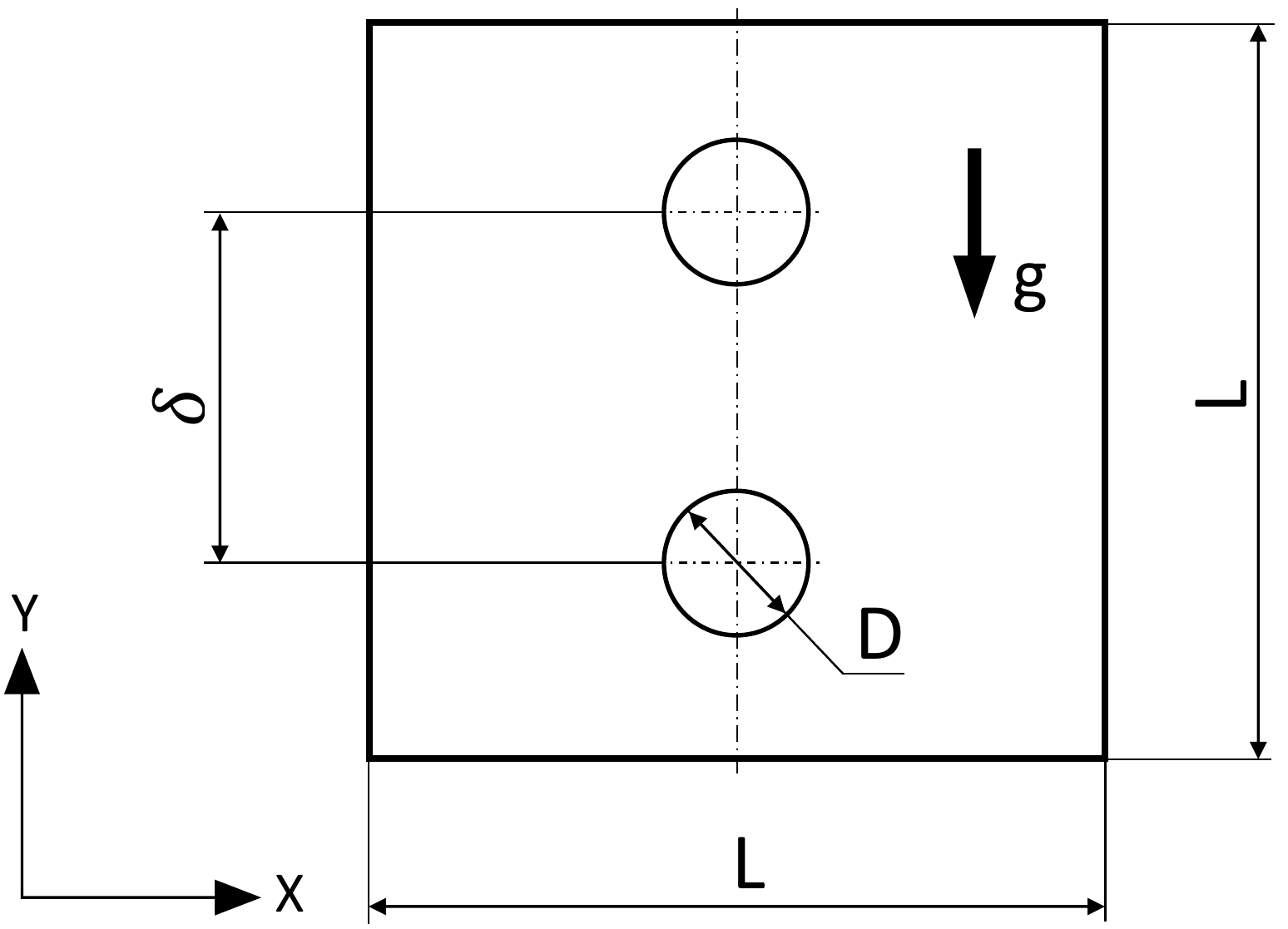}
\label{fig:VerTandemSchematic}
\end{figure}
\begin{figure}[h!]
\centering
   \subfigure[]
    {
        \includegraphics[width=0.47\textwidth,clip=]{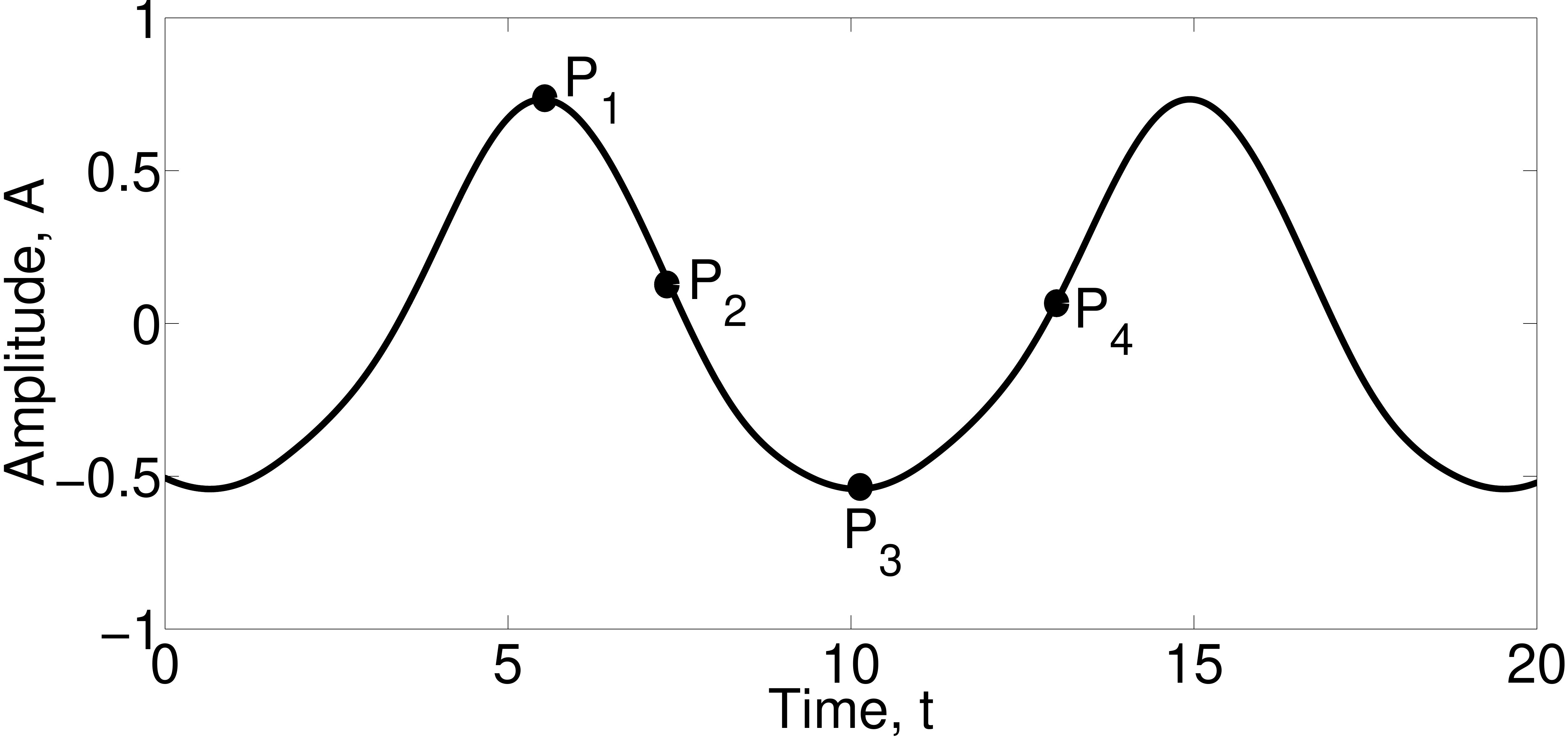}
    }
    \subfigure[]
     {
        \includegraphics[width=0.47\textwidth,clip=]{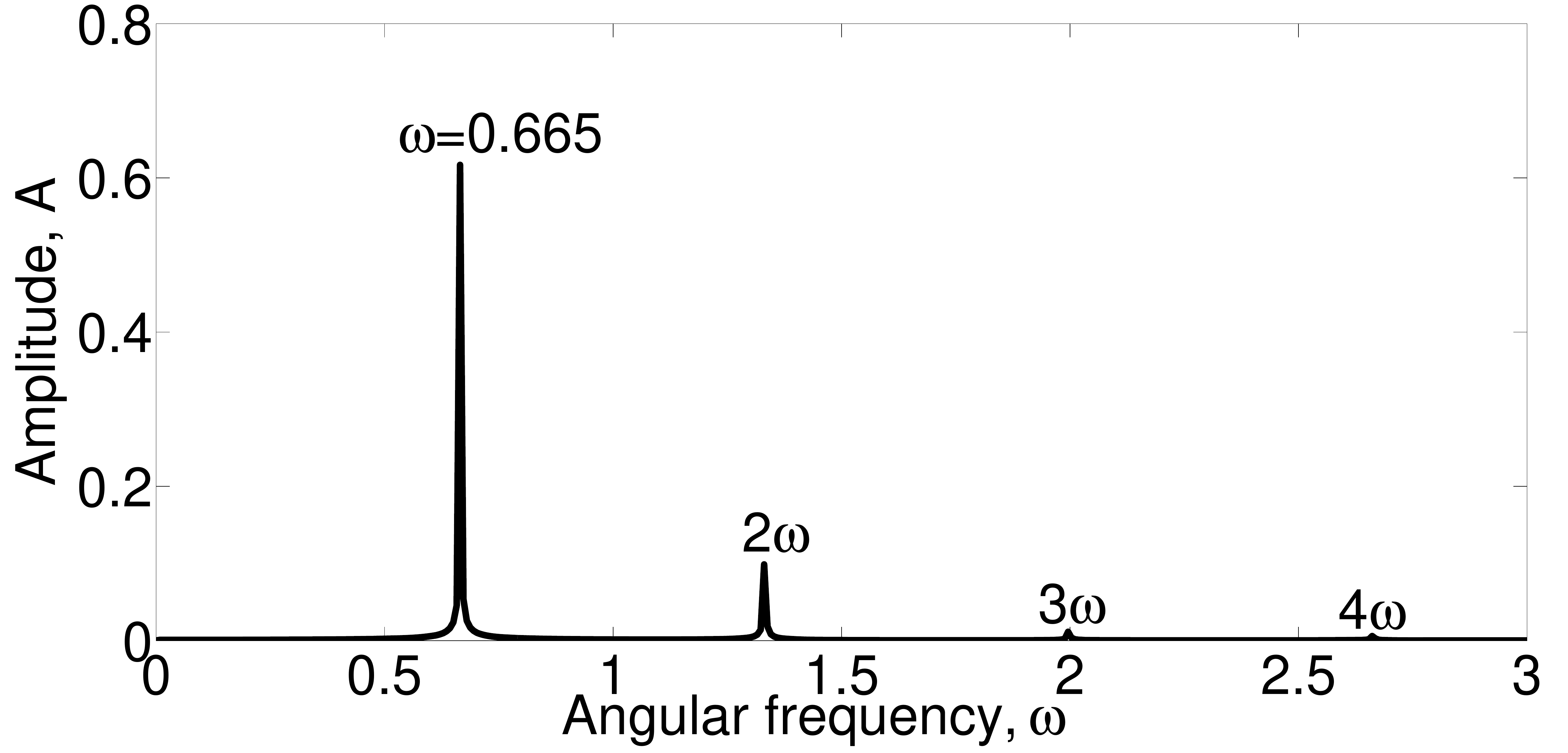}
    }

    \subfigure[]
    {
        \includegraphics[width=0.232\textwidth,clip=]{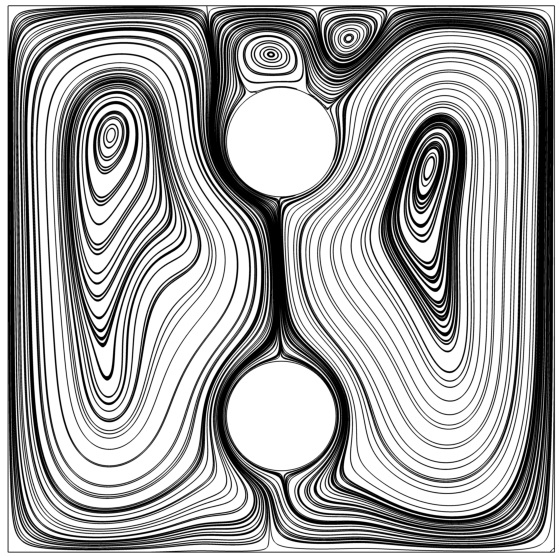}
        \includegraphics[width=0.23\textwidth,clip=]{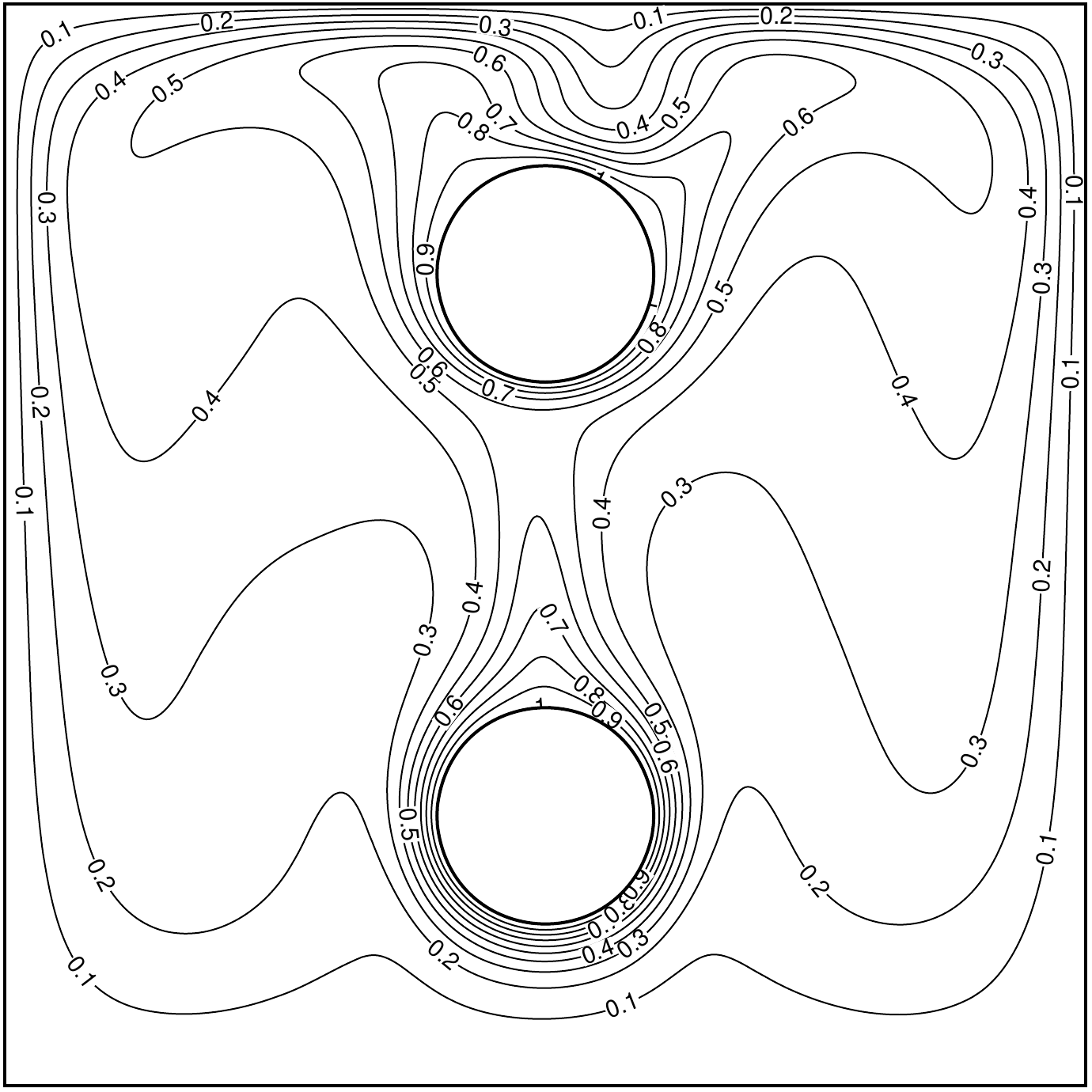}
    }
     \subfigure[]
    {
        \includegraphics[width=0.23\textwidth,clip=]{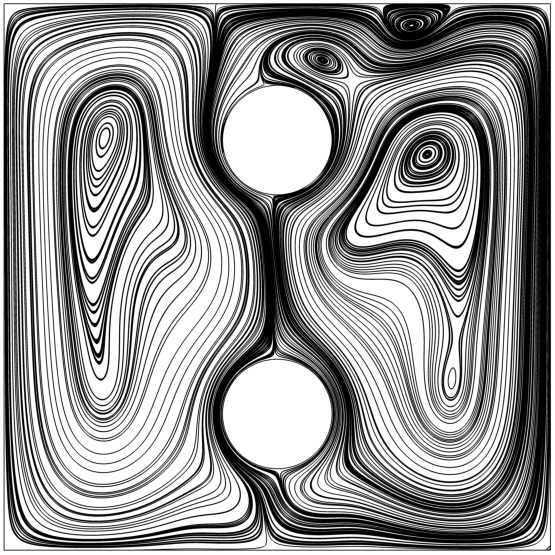}
        \includegraphics[width=0.23\textwidth,clip=]{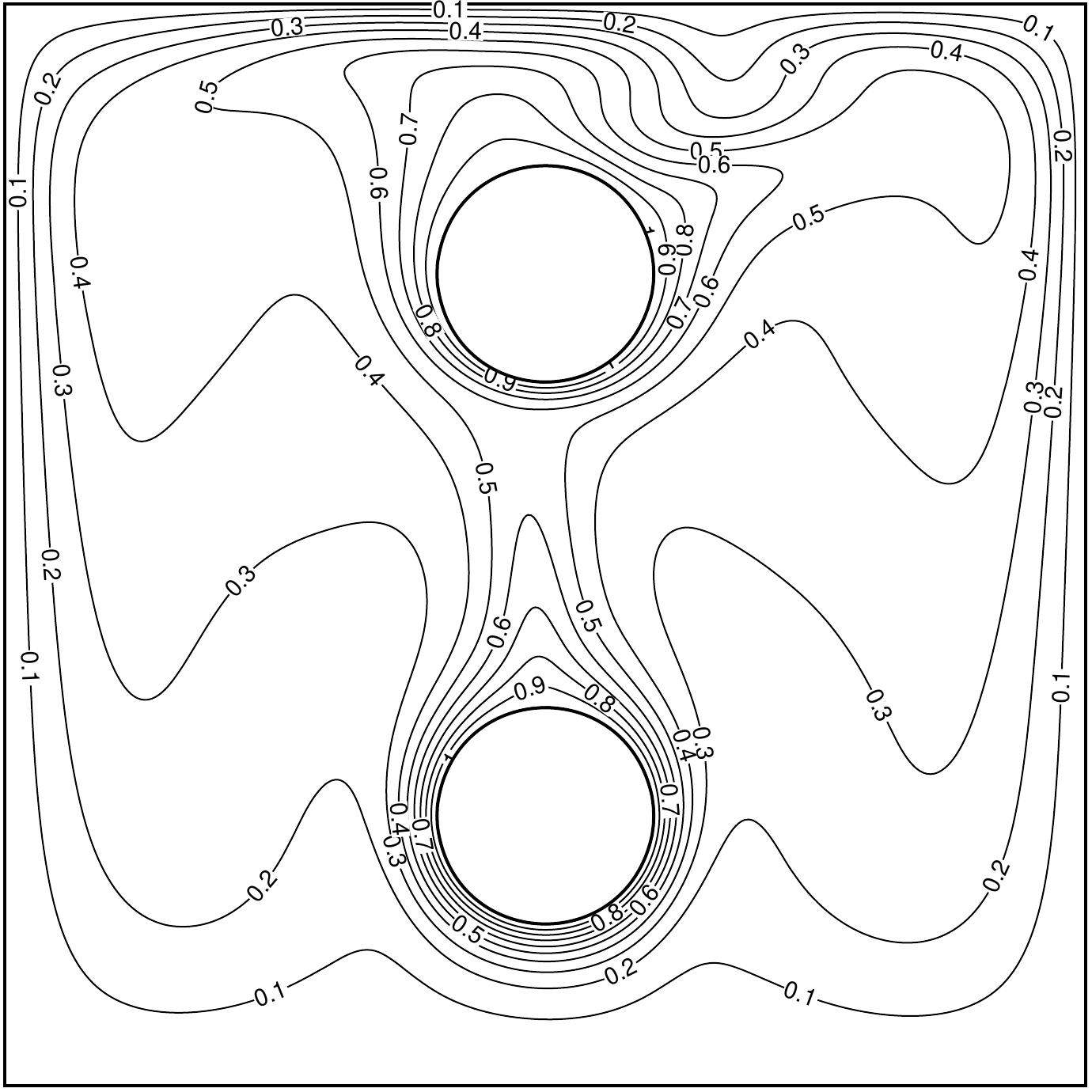}
    }

    \subfigure[]
    {
        \includegraphics[width=0.23\textwidth,clip=]{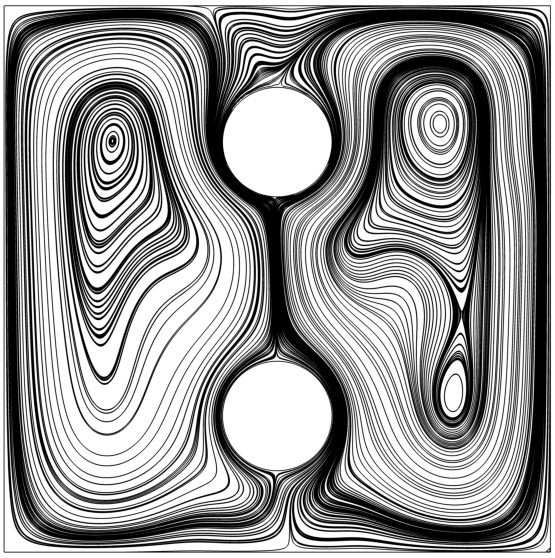}
        \includegraphics[width=0.23\textwidth,clip=]{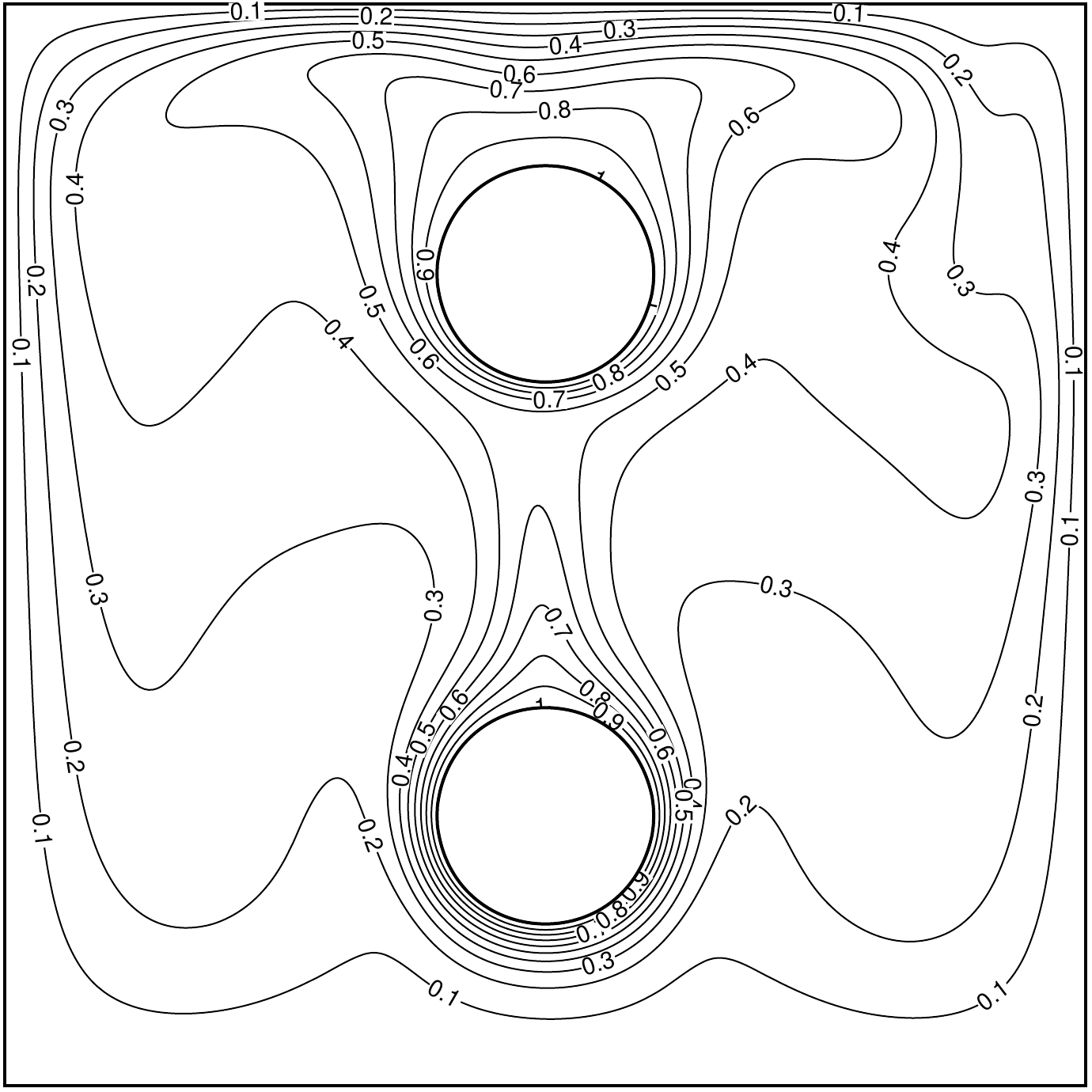}
    }
     \subfigure[]
     {
        \includegraphics[width=0.23\textwidth,clip=]{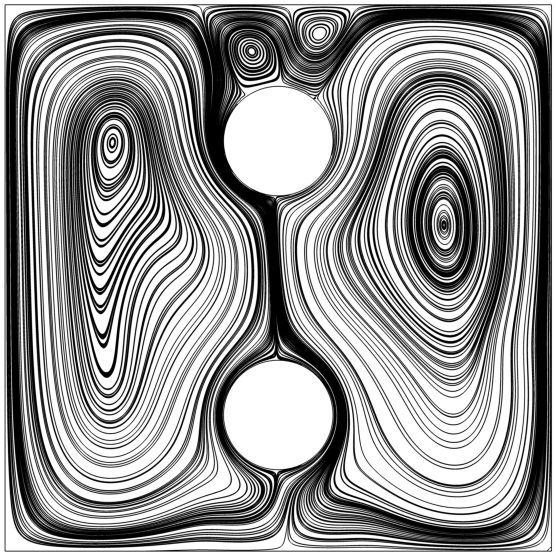}
        \includegraphics[width=0.23\textwidth,clip=]{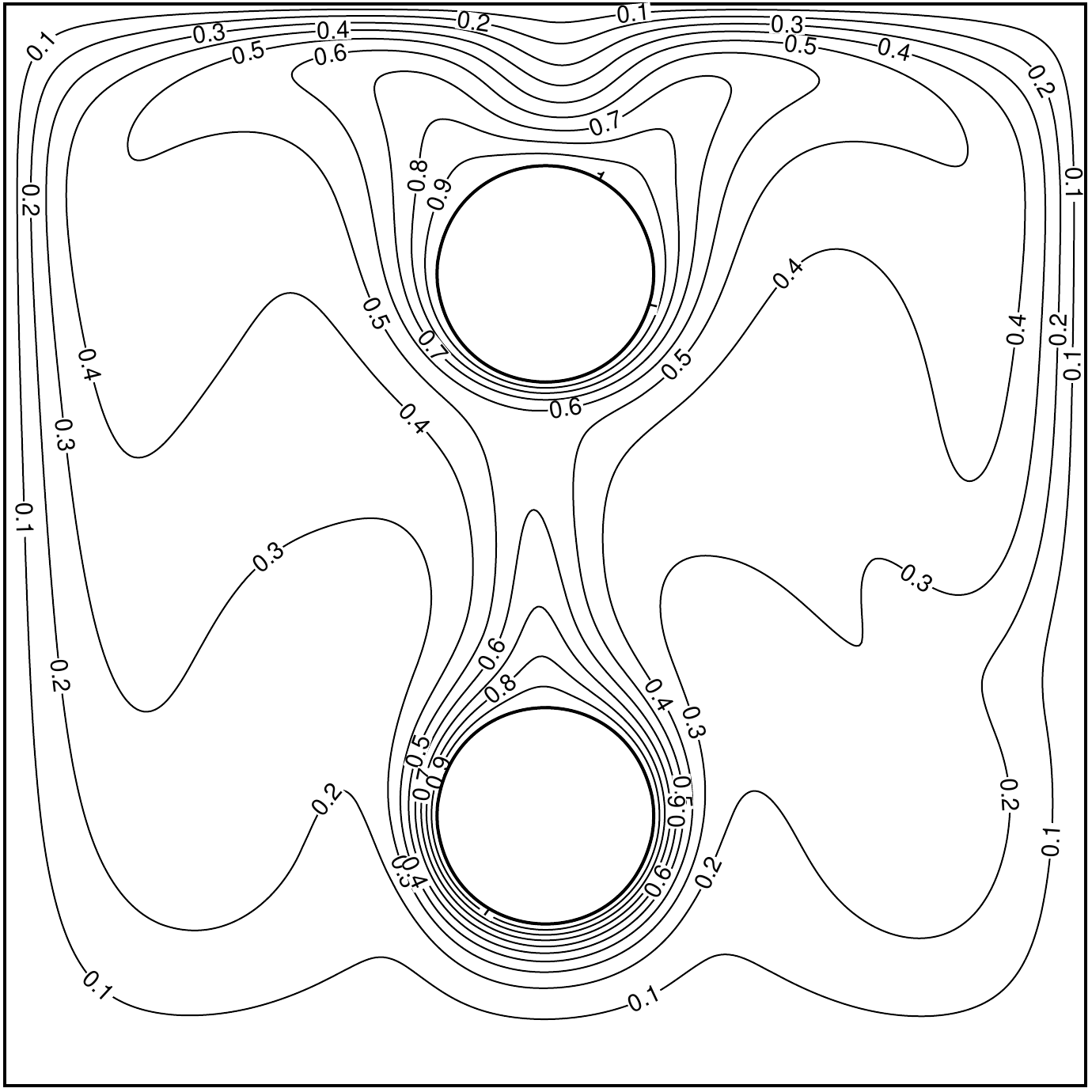}
    }
\caption{Characteristics of periodic natural convection flow developing around two vertically aligned cylinders located inside a square cavity with all cold boundaries at $Ra=10^6$  for $\delta=0.5$: (a) time evolution of the fluctuation, $A$, of the Nusselt number $\overline{Nu_{l}}$ averaged over the surface of the lower cylinder ($A= \overline{Nu_l}-mean(\overline{Nu_l}))$; (b) frequency spectrum of $A$; (c)-(f) instantaneous streamlines and isotherms at the selected time instances $[P_1, P_2, P_3, P_4]$.  }
\label{fig:ConvectTime}
\end{figure}
Figure \ref{fig:ConvectTime} presents the flow characteristics  of the periodic flow simulated at $Ra=10^6$. A grid independence study  was performed by comparing the velocity and temperature fields obtained on $400 \times 400$ and $500 \times 500$ grids. The maximum difference for all the flow characteristics obtained on the two grids did not exceed $0.5\%$, thereby successfully verifying the grid independence of the results. Figs. \ref{fig:ConvectTime}(a)-(b) present the time evolution of the amplitude, $A$ ( $A= \overline{Nu_l}-mean(\overline{Nu_l})$, where $mean(\overline{Nu_l})$ is the time averaged ($\overline{Nu_l}$) of the fluctuation of the average $\overline{Nu_l}$ number obtained for the lower cylinder and the corresponding frequency spectrum of $A$. Note that in agreement with the recent study of Park \cite{park2014} the flow at $Ra=10^6$ is governed by the single harmonics and its multipliers resulting from the flow non-linearity. The value of the angular frequency $\omega=0.665$ is in a good agreement with the corresponding reference value, $\omega_{ref}=0.656$,  reported in \cite{park2014}\footnote{Rescaled equivalent. Values reported in \cite{park2014} were multiplied by the factor $\frac{1}{Pr\sqrt{Gr}}$ to fit the sacaling adopted in this study.}. Instantaneous streamlines and isotherms shown in Figs. \ref{fig:ConvectTime} (c)-(e) for the four representative times $[P_1, P_2, P_3, P_4]$, (see Fig. \ref{fig:ConvectTime} (a)) evenly distributed over the single oscillating period are also in excellent agreement with the corresponding patterns reported in \cite{park2014}.

\subsection{Steady-state flow: steady incident flow around a circular cylinder}
This section presents a verification study of the fully implicit pressure-velocity coupled IB steady solver, based on the full Newton iteration, as defined by Eqs.(\ref{SteadyIB})-(\ref{NewtonIB2}).
The characteristics of the wake structure, typical of isothermal steady state flow around a circular cylinder are defined in Fig. \ref{fig:WakeSchema}. The steady flow was simulated at $Re=20$ and $Re=40$, and the obtained results were compared with previously published data. The simulations were performed utilizing the same computational set up (including the geometry and the boundary conditions) as  for the tandem arrangement of two cylinders (see the previous section) by omitting the back cylinder.
\begin{figure}[h!]
\includegraphics[width=0.6\textwidth,clip=]{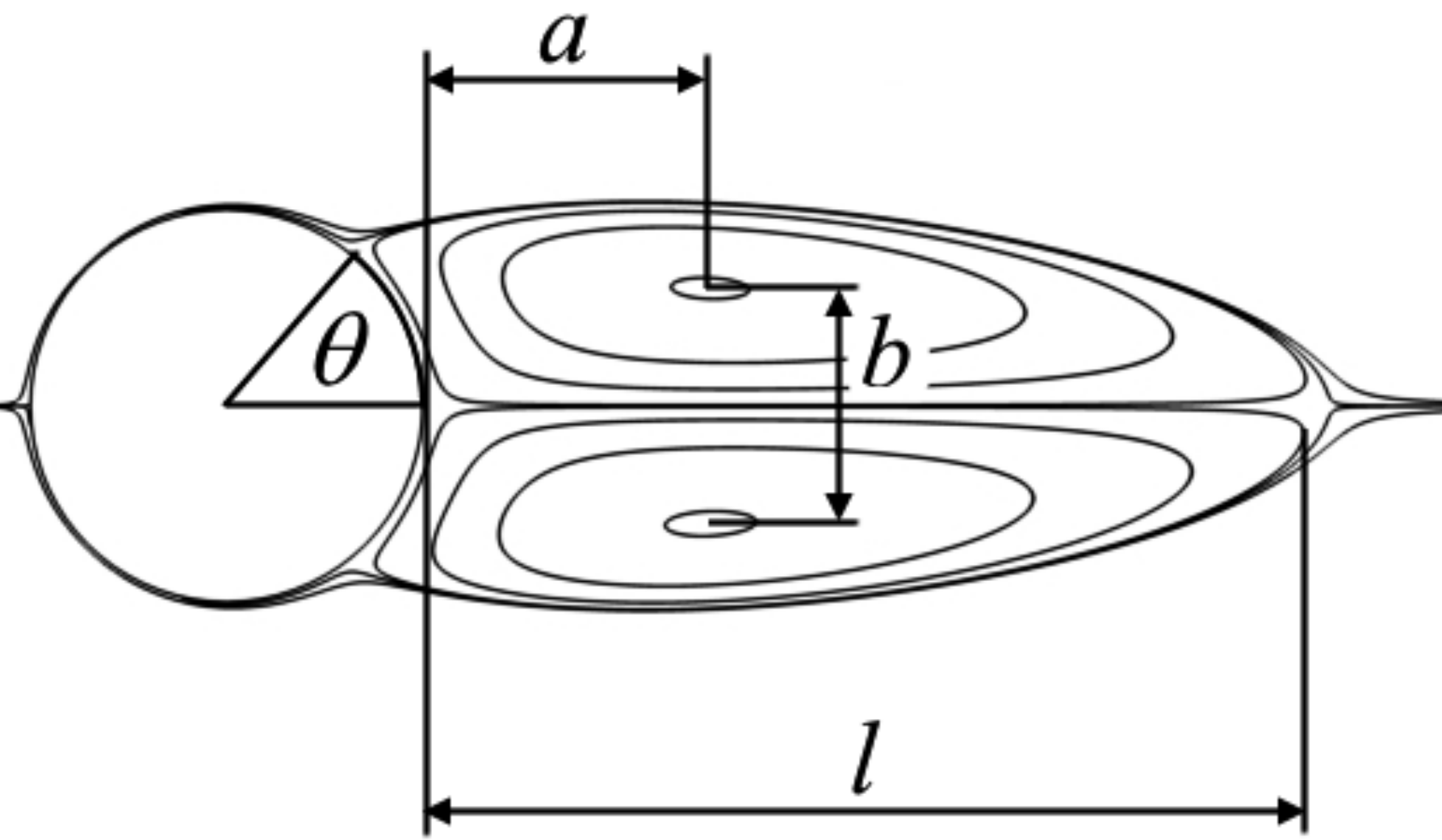}
\caption{Typical geometrical definitions of the steady state wake structure.}
\label{fig:WakeSchema}
\end{figure}
Figs. \ref{fig:cylSteady}(a) and (b) demonstrate the typical steady flow patterns developing around a horizontal cylinder at  $Re=20$ and $Re=40$, respectively. As expected, the flow is symmetric  relative to the horizontal centerline with two recirculating bubbles, located behind the cylinders.
\begin{figure}
\centering
    \subfigure[]
    {
        \includegraphics[width=0.45\textwidth,clip=]{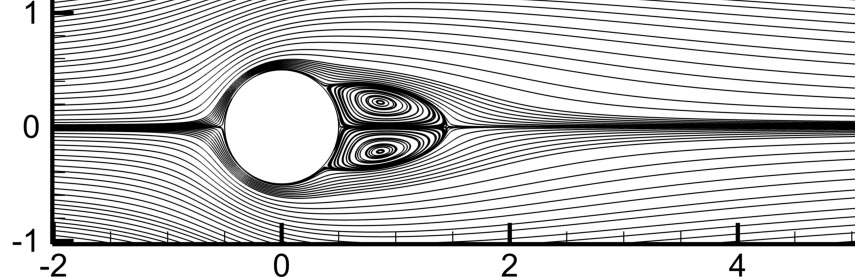}
    }
    \subfigure[]
    {
        \includegraphics[width=0.45\textwidth,clip=]{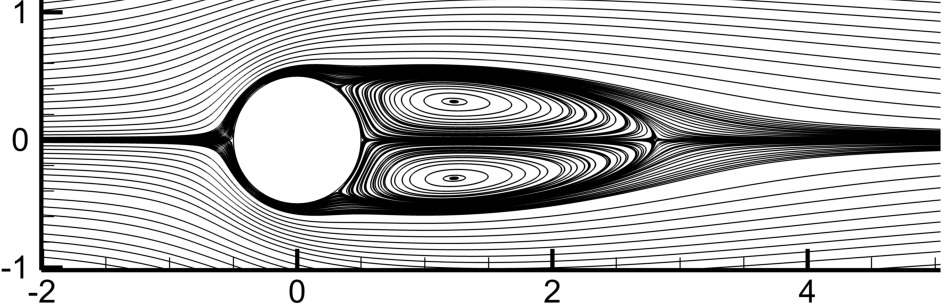}
    }

\caption{Streamlines of the steady-state flow around a circular cylinder obtained for: (a) $Re=20$; (b) $Re=40$  }
\label{fig:cylSteady}
\end{figure}
The geometrical characteristics of the bubbles for the two different values of $Re$ were compared with the literature data (see Table \ref{BaubleComparison}).
\begin{table}[]
\centering
\caption{Comparison of the wake characteristics and drag coefficients for steady-state flow over a cylinder for $Re = 20$ and $Re=40$. Experimental results are denoted by a $\star$ symbol.}
\label{BaubleComparison}
\begin{tabular}{llccccc}
\hline
        &                                                         &  $l/d$   &  $a/d$ & $b/d$ & $\theta$       &  $C_D$      \\ \hline
$Re=20$ &  Present                                                &  0.95    &  0.37  & 0.43  &  42.9\degree   &  2.09       \\
        &  Coutanceau and Bouard$^\star$ \cite{Coutanceau1977JFM} &  0.93    &  0.33  & 0.46  &  45.0\degree   &  --         \\
        &  Taira and Colonius \cite{taira2007JCP}                 &  0.94    &  0.37  & 0.43  &  43.3\degree   &  2.06       \\
        &  Linnick and Fasel \cite{linnick2005JCP}                &  0.93    &  0.36  & 0.43  &  43.5\degree   &  2.06       \\
        &  Dennis and Chung \cite{dennisJFM70}                    &  0.94    &   --   &  --   &  43.7\degree   &  2.05       \\
 $Re=40$&  Present                                                &  2.13    &  0.76  & 0.59  &  53.3\degree   &  1.56       \\
        &  Coutanceau and Bouard$^\star$ \cite{Coutanceau1977JFM} &  0.93    &  0.33  & 0.46  &  53.8\degree   &  --         \\
        &  Taira and Colonius \cite{taira2007JCP}                 &  2.30    &  0.73  & 0.60  &  53.7\degree   &  1.54       \\
        &  Linnick and Fasel  \cite{linnick2005JCP}               &  2.28    &  0.72  & 0.60  &  53.6\degree   &  1.54       \\
        &  Dennis and Chung \cite{dennisJFM70}                    &  2.35    &   --   & --    &  53.8\degree   &  1.52       \\\hline
\end{tabular}
\end{table}
Excellent quantitative agreement was observed between all the wake characteristics simulated in this study and those reported in the literature, thus verifying the developed steady-state solver for isothermal incompressible flows.

\subsection{Steady-state flow: steady natural convection confined flow}
Simulation of  natural convection confined flow was the focus of the next verification study. A configuration comprising an isothermal hot circular cylinder located at the center of a square cavity with all isothermal cold boundaries was chosen. Distributions of the isotherm contours and the stream function for three different $R/L$ ratios, $R/L=0.1, 0.2, 0.3$, as shown in Fig. \ref{fig:ConvctSteady}, were in excellent agreement with the corresponding data reported by Seta \cite{Seta2013physrev}.
\begin{figure}
\centering
    \subfigure[]
    {
        \includegraphics[width=0.325\textwidth,clip=]{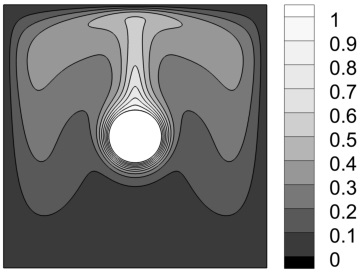}
        \includegraphics[width=0.33\textwidth,clip=]{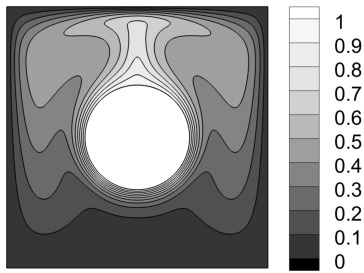}
        \includegraphics[width=0.33\textwidth,clip=]{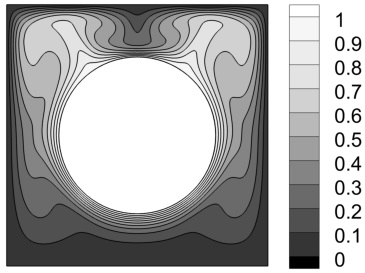}

    }
     \subfigure[]
    {   \includegraphics[width=0.335\textwidth,clip=]{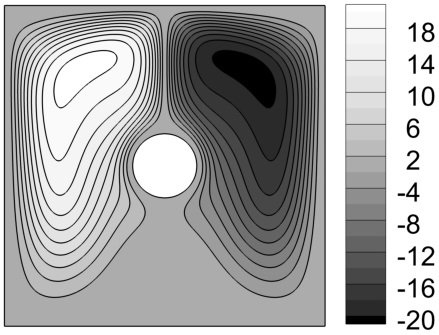}
        \includegraphics[width=0.322\textwidth,clip=]{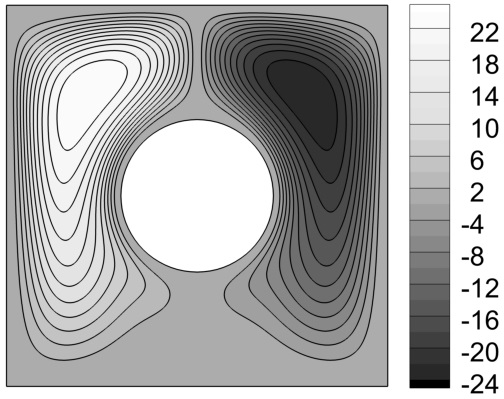}
        \includegraphics[width=0.327\textwidth,clip=]{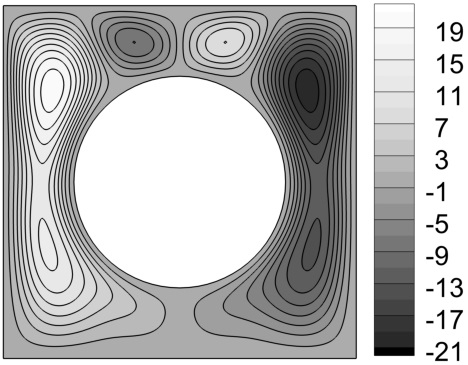}
    }
\caption{Flow characteristics, obtained for $Ra=10^6$, and $R/L=0.1, 0.2,0.3$: (a) isotherm contours; (b) stream function.}
\label{fig:ConvctSteady}
\end{figure}
Table \ref{RingComparison} presents a quantitative comparison of the $\overline{Nu}$ numbers, averaged over the cylinder surface, and of the absolute maximum values of the stream function $|\Psi_{max}|$ with the corresponding literature values. Acceptable agreement was found between our values and those reported in the literature for the entire range of $Ra$ and $R/L$ ratio values, thus verifying the developed steady-state solver applied to the simulation of natural convection flows.
\begin{table}[]
\centering
\caption{Comparison of the $\overline{Nu}$ number averaged over the cylinder surface and of the maximum absolute values of the stream function (rescaled equivalent, the calculated values of stream function were multiplied by the $Pr\sqrt{Gr}$ factor). All the results for the current study were obtained on a $400\times400$ grid.}
\label{RingComparison}
\begin{tabular}{llcccccc}
\hline
          &                                                         & \multicolumn{2}{c}{$Ra=10^4$}  & \multicolumn{2}{c}{$Ra=10^5$}  &  \multicolumn{2}{c}{$Ra=10^6$}       \\
          &                                                         &  $\overline{Nu}$  &$|\Psi_{max}|$  &  $\overline{Nu}$  &$|\Psi_{max}|$           & $\overline{Nu}$  &$|\Psi_{max}|$     \\\hline
$R/L=0.1$ &  Present                                                &     2.083           &  1.768            & 3.803 &          10.05  &              6.146  &            20.78     \\
          &  Seta \cite{Seta2013physrev}                            &     2.206           &  1.743            & 3.987 &          10.11  &              6.542  &            21.05     \\
          &  Moukalled and Acharya \cite{moukalled1996termopphys}   &     2.071           &  1.73             & 3.825 &          10.15  &              6.107  &            25.35     \\
          &  Shu and Zhu \cite{shu2002NemerFl}                      &     2.08            &  1.71             & 3.79  &          9.93   &              6.11   &            20.98     \\
 $R/L=0.2$&  Present                                                &     0.95            &  0.997            & 0.43  &          8.271  &              8.949  &            23.92     \\
          & Seta \cite{Seta2013physrev}                             &     3.461           &  0.981            & 5.253 &          8.267  &              9.547  &            24.23     \\
          &  Moukalled and Acharya \cite{moukalled1996termopphys}   &     3.331           &  1.02             & 5.08  &          8.38   &              9.374  &            24.07     \\
          &  Shu and Zhu \cite{shu2002NemerFl}                      &     3.24            &  0.97             & 4.86  &          8.10   &              8.90   &            24.13     \\
$R/L=0.3$ &  Present                                                &     5.402           &  0.494            & 6.246 &          5.046  &              11.967 &            20.23     \\
          &  Seta \cite{Seta2013physrev}                            &     5.832           &  0.486            & 6.685 &          5.023  &              12.87  &            20.33     \\
          &  Moukalled and Acharya \cite{moukalled1996termopphys}   &     5.826           &  0.50             & 6.107 &          5.10   &              11.62  &            21.30     \\
          &  Shu and Zhu \cite{shu2002NemerFl}                      &     5.40            &  0.49             & 6.21  &          5.10   &              12.00  &            20.46     \\\hline

\end{tabular}
\end{table}
\subsection{Linear stability analysis: incident flow around a circular cylinder}
The developed IB FPCD linear stability solver was verified by conducting a linear stability analysis of the incident base flow around a circular cylinder. The base flow was calculated by the steady-state solver, which was verified in the previous section. The same computational domain and grid resolution as  for the case of steady-state flow analysis around a circular cylinder were utilized. Perturbation values of all the flow fields were set to zero at all the boundaries. Figure \ref{fig:CylinerStabVerif}-a and \ref{fig:CylinerStabVerif}-b present the contours of vorticity of the real and imaginary parts of the leading eigenmode, respectively. Both patterns are symmetric relative to the $Y=0$ line and are characterized by alternating minima and maxima values in the streamwise direction. Contours of the vorticity of the imaginary part of the leading eigenmode (Fig. \ref{fig:CylinerStabVerif}-b) bear a striking resemblance to the corresponding pattern reported by Barkley \cite{barkley2006}. Figures \ref{fig:CylinerStabVerif}-c and \ref{fig:CylinerStabVerif}-d present a quantitative comparison between the obtained and reference \cite{barkley2006} values for the frequency and growth rate, calculated in the range of $30\leq Re\leq 180$ by the linear stability analysis. It can be seen that both quantities are in excellent agreement for the entire range of $Re$ values, thereby successfully verifying the developed linear stability solver.
\begin{figure}
\centering
    \subfigure[]
    {
        \includegraphics[width=0.45\textwidth,clip=]{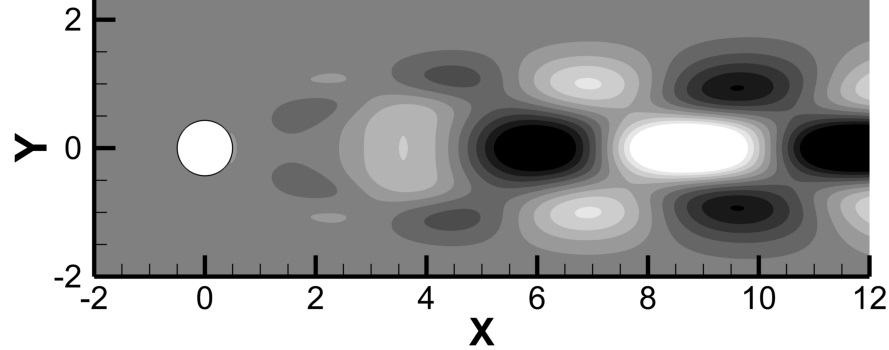}
    }
     \subfigure[]
    {
        \includegraphics[width=0.45\textwidth,clip=]{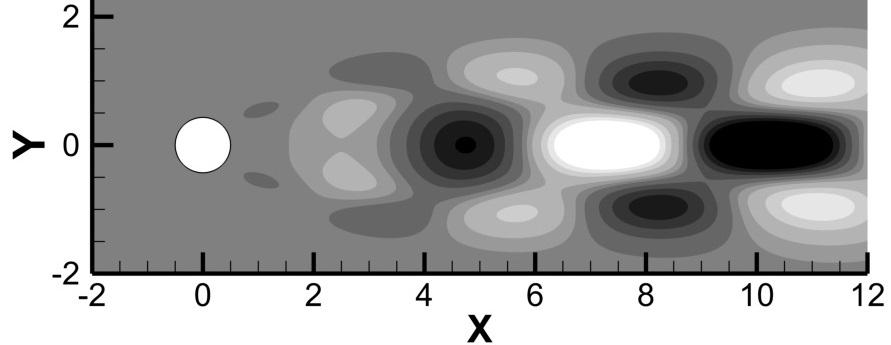}
    }
    \subfigure[]
    {
        \includegraphics[width=0.5\textwidth,clip=]{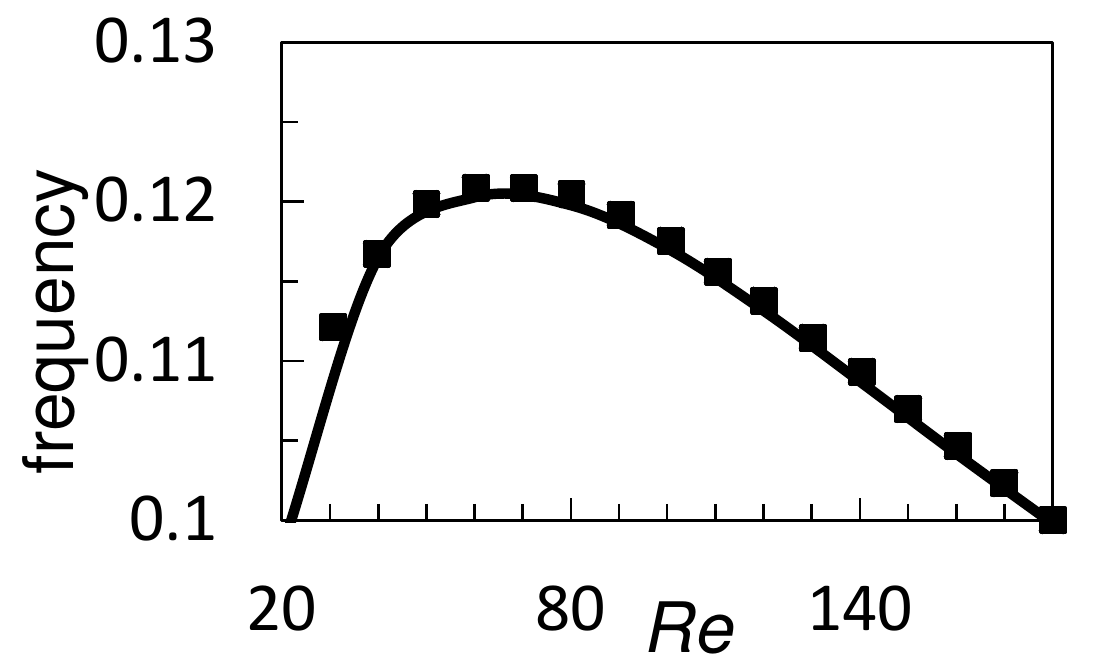}
    }
    \subfigure[]
    {
        \includegraphics[width=0.45\textwidth,clip=]{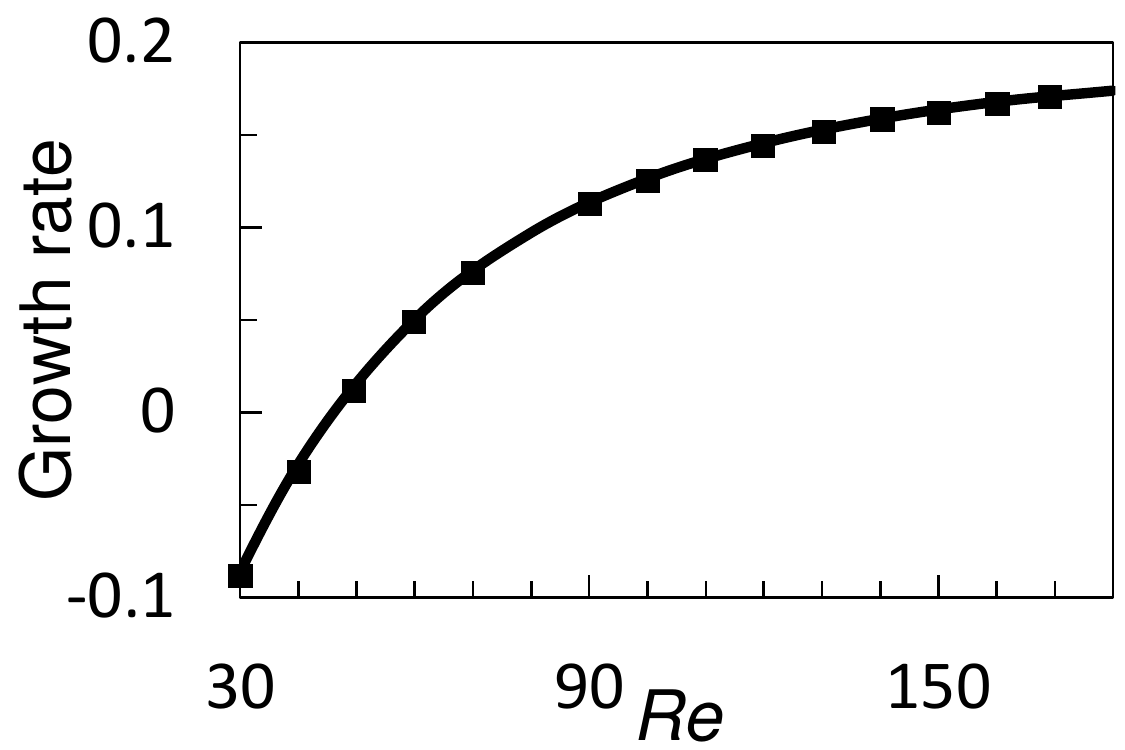}
    }
\caption{Results of the linear stability analysis of incident flow over the cylinder at $Re=100$: (a) real part of the leading eigenmode of  vorticity; (b) imaginary part of the leading eigenmode of vorticity; (c) comparison of  the critical frequency values (bold line) obtained in this study with the reference values (black squares, scanned from \cite{barkley2006}) as a function of the $Re$ number; (d) comparison of the growth rate values (bold line)obtained in this study  with the reference values (black squares, scanned from \cite{barkley2006}) as a function of the $Re$ number.}
\label{fig:CylinerStabVerif}
\end{figure}

\subsection{Linear stability analysis: incident flow around two horizontally aligned circular cylinders}
Figure \ref{fig:TandemGenStab}-a presents a quantitative comparison of the critical values of the $Re$ number in this study obtained by the linear stability analysis with the corresponding reference values reported by Carmo et al. \cite{carmo2010PhysFl} for the flow around two horizontally aligned cylinders. The simulations were performed by utilizing the same setup (size of the computational domain and grid resolution) as for the analysis of unsteady flow, detailed in Section \ref{UnstableTandem}. Good agreement was obtained between the values calculated in this study and the reference $Re_{cr}$ values for the entire range of non-dimensional distances between the cylinder centers, $l/d$. Note the non-homogeneity of the $Re-l/d$ functionality for the investigated range of $l/d$ values. The non-homogeneity can be explained by the existence of three  different vortex shading regimes SG, AG, and WG. It is noteworthy that the $f_{cr}-l/d$ functionality, shown in Fig. \ref{fig:TandemGenStab}-b, exhibits a different trend. The value decays continuously for the range of $1.5 \leq l/d < 4$.  Thereafter, the trend is reversed, and the value of $f_{cr}$ grows continuously, attaining a maximum at $l/d\approx4.9$. Finally, a rapid decrease of the $f_{cr}$ value is observed at $l/d=5$.
\begin{figure}
\centering
    \subfigure[]
    {
        \includegraphics[width=0.45\textwidth,clip=]{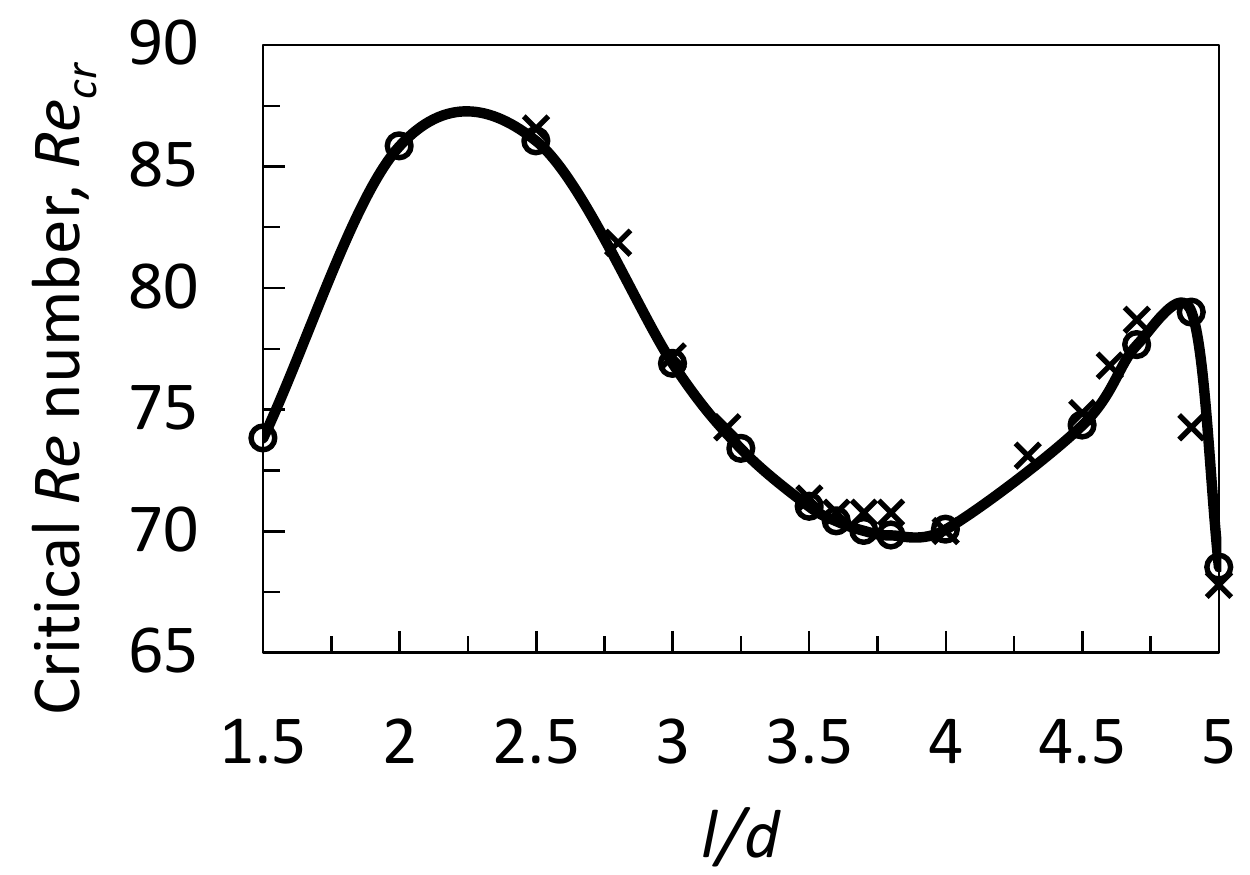}
    }
    \subfigure[]
    {
        \includegraphics[width=0.45\textwidth,clip=]{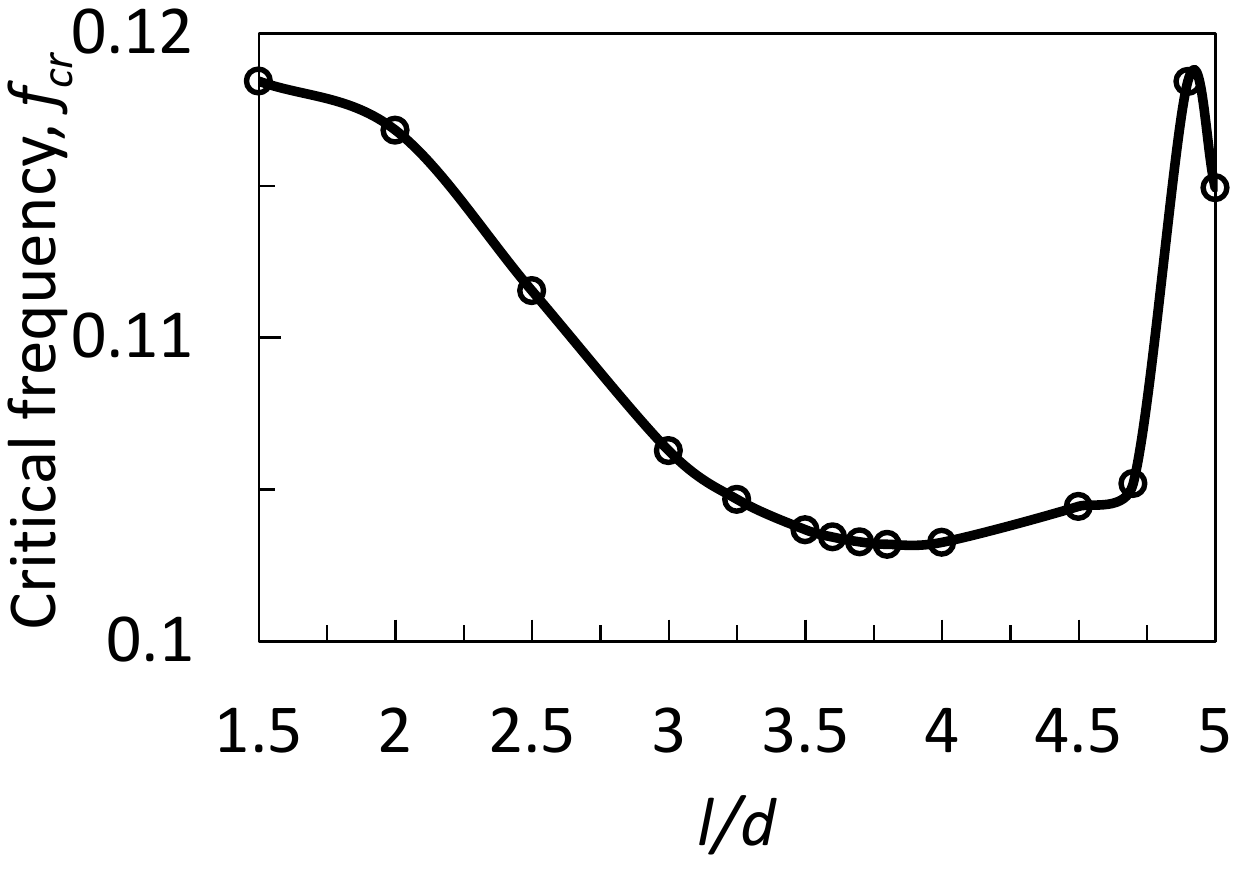}
    }

\caption{Values of critical Reynolds number, $Re_cr$ and critical frequency, $f_{cr}$ versus non-dimensional distance, $l/d$, obtained for the flow around two circular cylinders in tandem. The $\times$ marker reads for the reference values of $Re_{cr}$ number as scanned from \cite{carmo2010PhysFl}.}
\label{fig:TandemGenStab}
\end{figure}

Additional evidence for the  existence of three different vortex shading regimes for the incident flow around two horizontally aligned cylinders as a function of the distance between the cylinder centers is provided by the contours of the corresponding leading eigenvectors obtained for $Re=Re_{cr}$. In fact, both the real and imaginary parts of the leading eigenvectors of vorticity exhibit different patterns for the three different values of $l/d$ distance, as shown in Fig. \ref{fig:tandemStability}. As expected, the largest differences in the perturbation fields are observed in the intermediate region between the cylinders and in the wake in the close vicinity of the trailing cylinder. Further away from the trailing cylinder, all the patterns corresponding to the leading eigenvectors of vorticity attain similar sign-alternating petal structures, symmetrically aligned along the horizontal centerline.

\begin{figure}
\centering
    \subfigure[]
    {
        \includegraphics[width=0.5\textwidth,clip=]{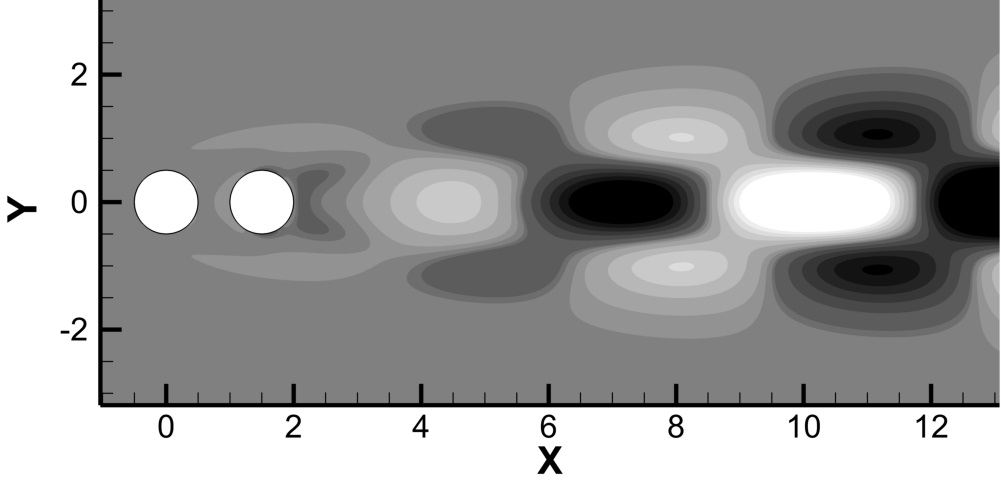}
        \includegraphics[width=0.5\textwidth,clip=]{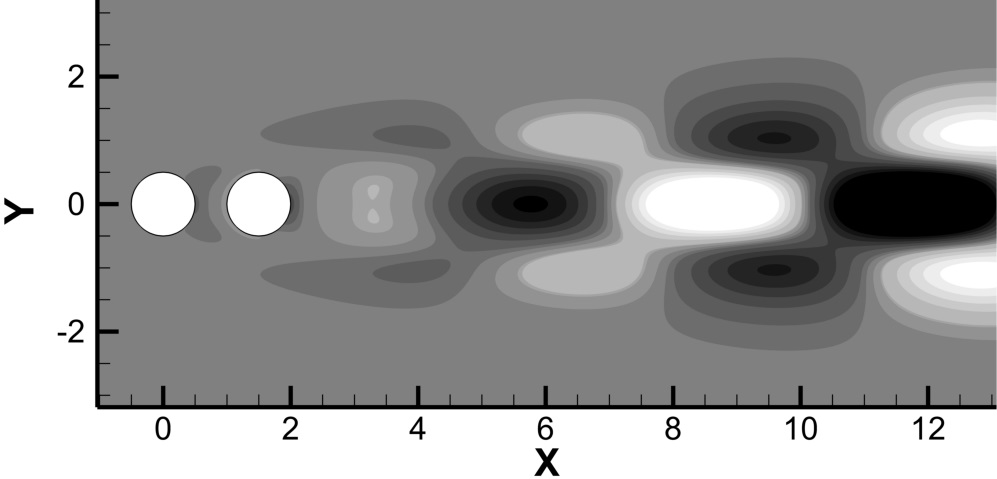}
    }

    \subfigure[]
    {
        \includegraphics[width=0.5\textwidth,clip=]{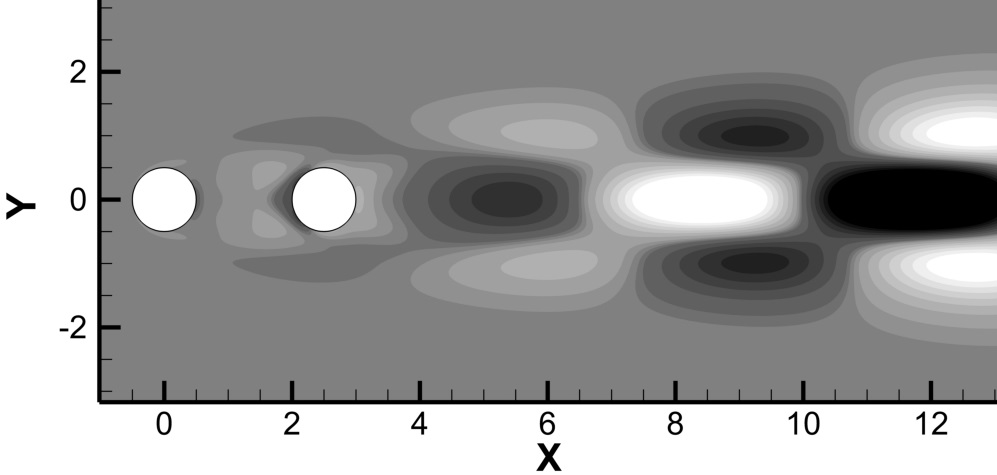}
        \includegraphics[width=0.5\textwidth,clip=]{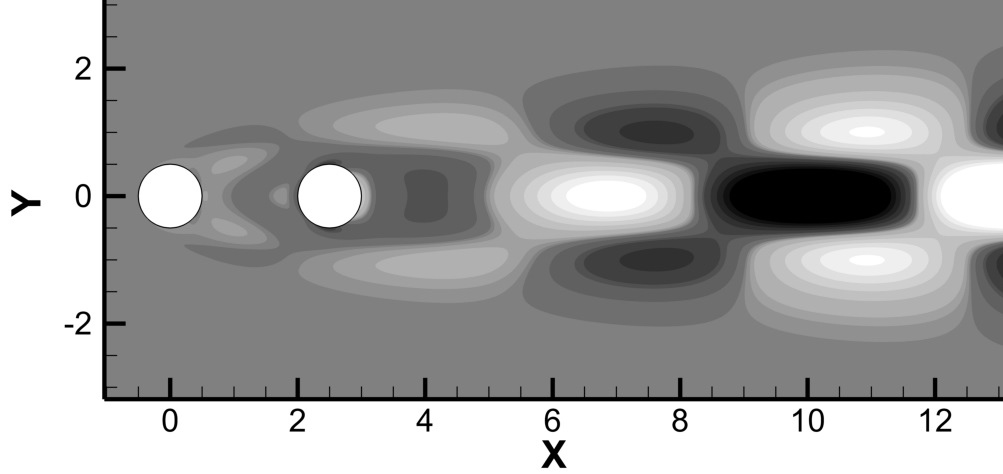}
    }

    \subfigure[]
    {
        \includegraphics[width=0.5\textwidth,clip=]{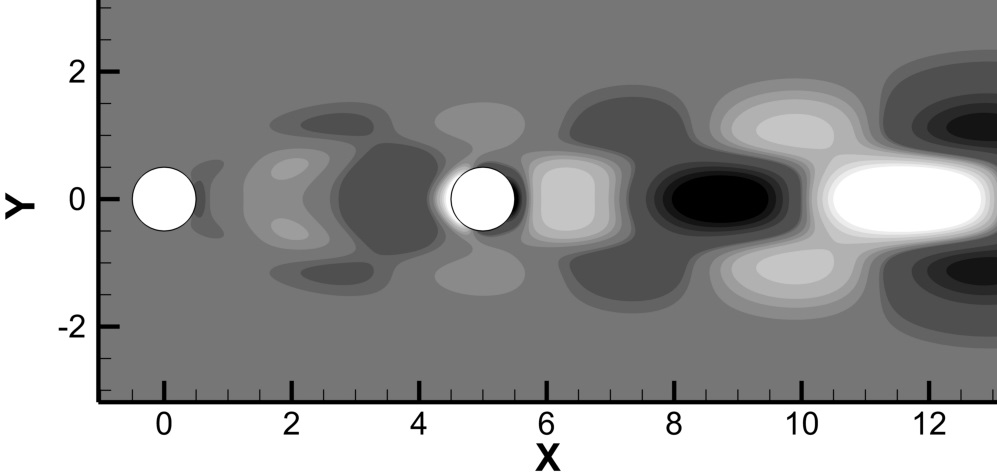}
        \includegraphics[width=0.5\textwidth,clip=]{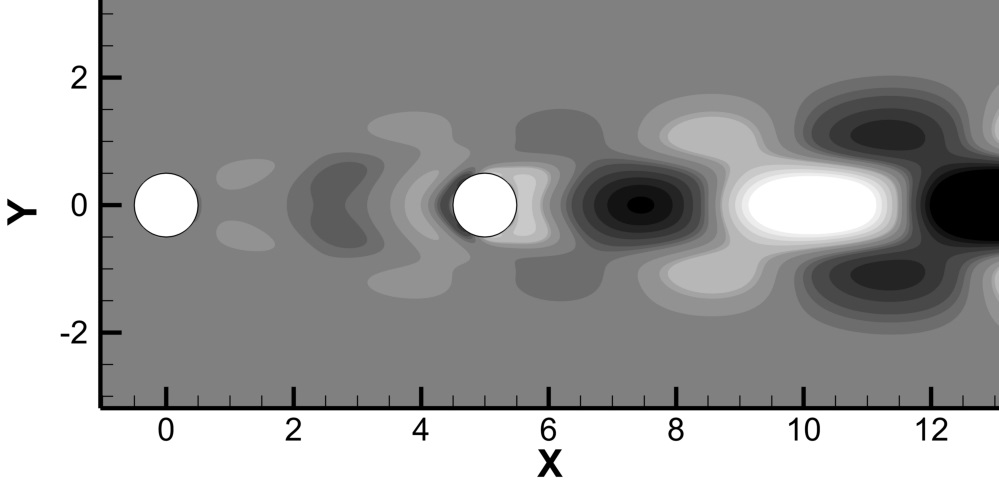}
    }

\caption{Results of a linear stability analysis of the incident flow over the two horizontally aligned cylinders at $Re=Re_{cr}$. Left and right columns correspond to the contours of  real and imaginary  parts of  vorticity of the leading eigenmode, respectively, for: (a) $l/d=1.5$ and $Re_{cr}=73.1$; (b) $l/d=2.5$ and $Re_{cr}=86.86$; (c) $l/d=5$ and $Re_{cr}=68.51$;}
\label{fig:tandemStability}
\end{figure}

\subsection{Linear stability analysis: confined natural convection flow}
This section focuses on the linear stability analysis of the confined natural convection flow around two hot vertically aligned circular cylinders located inside a square cavity with all cold boundaries. For this configuration, unsteady periodic flow was observed at $Ra=10^6$ (see section \ref{UnsteadyNatConvect}), thereby providing the lower value of $Ra_{cr}$ for the first Hopf bifurcation. The contours of the absolute values of the leading eigenmode of the temperature $|\theta'|$ and the kinetic energy $|e'|=0.5(|u'|^2+|v'|^2)$ obtained at $Ra_{cr}=5.026\times 10^5$ on a $800 \times 800$ grid are shown in Fig. \ref{fig:NatConvect2CylindNatSatab}. For both quantities, the region characterized by the highest amplitude of perturbation can be clearly recognized immediately above the top cylinder. Both amplitude distributions are not symmetric and are biased to the right (but could also be biased to the left with the same probability) relative to the vertical centerline. The obtained patterns are consistent with the structure of supercritical flow at $Ra=10^6$  (see Fig. \ref{fig:ConvectTime}), clearly indicating the interaction between two counter rotating vortices as the primary source of the observed instability.
\begin{figure}
\centering
    \subfigure[]
    {
        \includegraphics[width=0.45\textwidth,clip=]{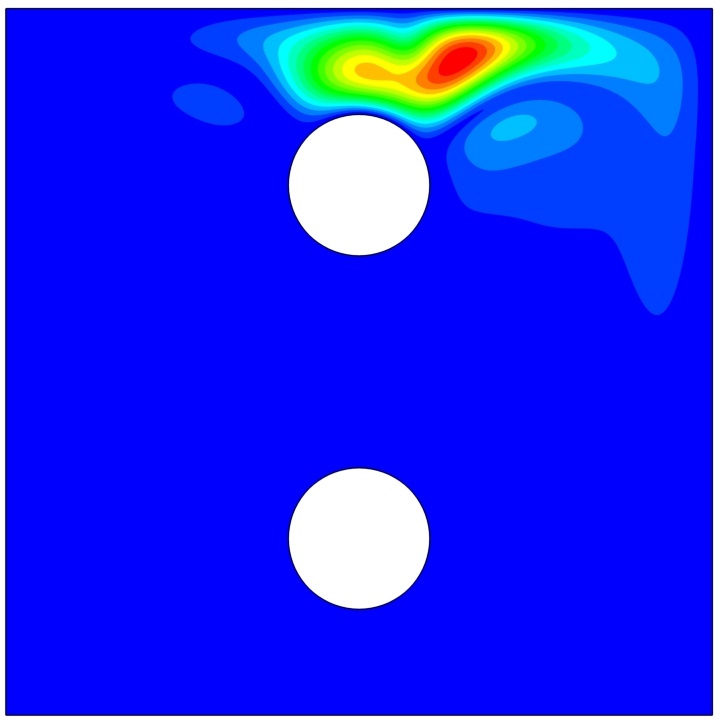}
    }
    \subfigure[]
    {
        \includegraphics[width=0.45\textwidth,clip=]{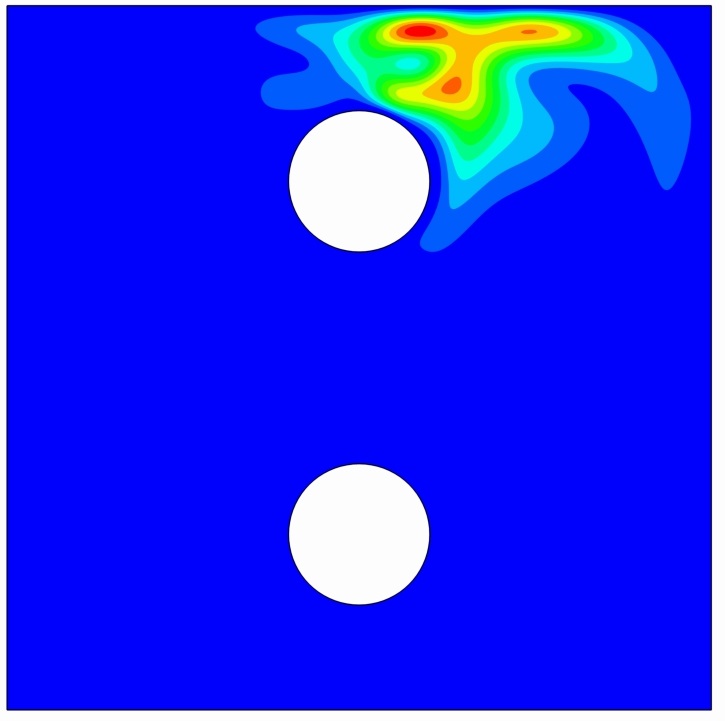}
    }

\caption{ Contours of absolute values of the leading eigenmode obtained at $Ra_{cr}=5.026\times 10^5$ and $\delta=0.5$ on a $800 \times 800$ grid for: (a)  temperature, $|\theta'|$ ; (b) kinetic energy, $|e'|=0.5(|u'|^2+|v'|^2)$. }
\label{fig:NatConvect2CylindNatSatab}
\end{figure}
A summary of the grid convergence study with respect to the obtained values of the critical Rayleigh number, $Ra_{cr}$, and the oscillating frequency $\omega_{cr}$ is presented in Table \ref{GridConvergence}. It can be seen that for the  $800 \times 800 $ grid the results have converged up to the third decimal digit, thereby verifying the independence of the obtained  $Ra_{cr}$ and $\omega_{cr}$ values on the resolution of the grid.
\begin{table}[]
\centering
\caption{Grid convergence for the critical  $Ra_{cr}$ and $\omega_{cr}$ values, $\delta=0.5$. }
\label{GridConvergence}
\begin{tabular}{ccc}
\hline
    Grid              &$Ra_{cr}$               &$\omega_{cr}$       \\
\hline
    300$\times$300    &5.074 $\times 10^5$     &0.2801              \\
    400$\times$400    &5.073 $\times 10^5$     &0.2816              \\
    500$\times$500    &5.041 $\times 10^5$     &0.2829              \\
    600$\times$600    &5.035 $\times 10^5$     &0.2842              \\
    700$\times$700    &5.030 $\times 10^5$     &0.2851              \\
    800$\times$800    &5.026 $\times 10^5$     &0.2855              \\
\hline
\end{tabular}
\end{table}
\newpage
\section{Summary and conclusions}
A new formulation of the IB method allowing for simulation of unsteady incompressible flows around immersed bodies of various shapes and for performing linear stability analysis of the flows was developed. The developed method is based on the fully pressure-velocity coupled approach, which implicitly provides a divergence-free velocity field,  with no need for an extra projection-correction step. The method treats pressure, boundary forces,  and heat sources as Lagrange multipliers, thereby implicitly satisfying the kinematic constraints of no-slip and the determined temperature (or heat source) boundary conditions. The developed method facilitated an efficient and highly accurate linear stability analysis of the incompressible flows in the presence of a variety of arbitrarily oriented immersed bodies of various shapes. The method was extensively verified for both isothermal and natural convection flows. The independence of the obtained results on the resolution of the computational grid was established by the excellent quantitative agreement with independent numerical and experimental results available in the literature and from the current convergence study.

Although the developed methodology is focussed on $2D$ incompressible flows, its extension to $3D$ is straight forward and is restricted only by the limitations of available computer memory. This new approach for the time integration and linear stability analysis of $3D$ flows will become more and more attractive as modern efficient direct solvers are developed and the power of computational resources increases. The new methodology can be also efficiently applied to the mesoscale linear stability analysis of quasi $2D$ flows in porous media,  which will be the focus of our future work.

\bibliographystyle{model1-num-names}
\bibliography{paper}







\end{document}